\documentclass[useAMS,usenatbib]{mnras}
\usepackage{newtxtext,newtxmath}
\usepackage[T1]{fontenc}

\DeclareRobustCommand{\VAN}[3]{#2}
\let\VANthebibliography\thebibliography
\def\thebibliography{\DeclareRobustCommand{\VAN}[3]{##3}\VANthebibliography}

\usepackage{graphicx}	
\usepackage{amsmath}	
\usepackage{listings}
\usepackage{siunitx}
\usepackage{pifont}
\usepackage{tikz}
\usepackage{subcaption}
\usepackage{hyperref}
\usepackage{lipsum}
\usepackage{acro}
\usepackage{physics}
\usepackage[capitalize]{cleveref}

\newcommand{\cmark}{\ding{51}}%
\newcommand{\xmark}{\ding{55}}%

\newcommand{\ba}{\[\begin{aligned}}
\newcommand{\ea}{\end{aligned}\]}
\usepackage{colortbl}
\newcommand{\eq}[1]{\begin{align}#1\end{align}}

\makeatletter
\g@addto@macro\bfseries{\boldmath}
\makeatother

\DeclareSIUnit \parsec {pc}
\DeclareSIUnit \h {h}
\DeclareSIUnit\year{yr}

\newcommand{\sphinx}{{\small SPHINX$^{20}$}}
\newcommand{\jades}{JADES}
\newcommand{\fesc}{$f_{\rm esc}$}

\newcommand{\xion}{${\xi_{\rm ion}}$}
\newcommand{\rhouv}{${\rho_{\rm UV}}$}
\newcommand{\ltuili}{{\small LTU-ILI}}
\newcommand{\nion}{$\dot{n}_{\rm ion}$}
\newcommand{\Nion}{$\dot{N}_{\rm ion}$}

\DeclareAcronym{AGN}{short = AGN, long  = active galactic nucleus}
\DeclareAcronym{SED}{short = SED, long  = spectral energy distribution}
\DeclareAcronym{ISM}{short = ISM, long  = interstellar medium}
\DeclareAcronym{IGM}{short = IGM, long  = intergalactic medium}
\DeclareAcronym{ILI}{short = ILI, long  = implicit likelihood inference}
\DeclareAcronym{SBI}{short = SBI, long  = simulation-based inference}
\DeclareAcronym{LFI}{short = LFI, long  = likelihood-free inference}
\DeclareAcronym{JWST}{short = JWST, long  = James Webb Space Telescope}
\DeclareAcronym{PIT}{short = PIT, long  = Probability Integral Transform}
\DeclareAcronym{ET}{short = ET, long  = Extra-Trees}
\DeclareAcronym{MAE}{short = MAE, long  = median absolute error}
\DeclareAcronym{IMF}{short = IMF, long  = initial mass function}

\title[The Ionizing Luminosities of EoR Galaxies]{Inferring the Ionizing Photon Contributions of High-Redshift Galaxies to Reionization with JWST NIRCam Photometry}

\author[N. Choustikov, et al.]{Nicholas Choustikov$^{1}$\thanks{nicholas.choustikov@physics.ox.ac.uk}, Richard Stiskalek$^{1}$, Aayush Saxena$^{1,2}$, Harley Katz$^{3}$, Julien Devriendt$^{1}$,\newauthor
and Adrianne Slyz$^{1}$
\\
$^{1}$Sub-department of Astrophysics, University of Oxford, Keble Road, Oxford OX1 3RH, United Kingdom\\
$^{2}$Department of Physics and Astronomy, University College London, Gower Street, London WC1E 6BT, United Kingdom\\
$^{3}$Department of Astronomy \& Astrophysics, University of Chicago, 5640 S Ellis Avenue, Chicago, IL 60637, USA\\
}

\date{Accepted XXX. Received YYY; in original form ZZZ}

\pubyear{2025}

\begin{document}
\label{firstpage}
\pagerange{\pageref{firstpage}--\pageref{lastpage}}
\maketitle

\begin{abstract}
\textit{JWST} observations are providing unprecedented constraints on the history of reionization owing to the ability to detect faint galaxies at $z\gg6$. Modeling this history requires understanding both the ionizing photon production rate ($\xi_{\rm ion}$) and the fraction of those photons that escape into the intergalactic medium ($f_{\rm esc}$). Observational estimates of these quantities generally rely on spectroscopy for which large samples with well-defined selection functions remain limited. To overcome this challenge, we present and release a novel implicit likelihood inference pipeline, \texttt{PHOTONIOn}, trained on mock photometry to predict the escaped ionizing luminosity of individual galaxies ($\dot{N}_{\rm ion}$) based on photometric magnitudes and redshifts. We show that \texttt{PHOTONIOn} is able to reliably infer $\dot{N}_{\rm ion}$ from photometry. This is in contrast to traditional SED-fitting approaches which rely on $f_{\rm esc}$ prescriptions that often over-predict $\dot{N}_{\rm ion}$ for LyC-dim galaxies, even when given access to spectroscopic data. We have deployed \texttt{PHOTONIOn} on a sample of 4,559 high-redshift galaxies from the JADES Deep survey, finding gentle redshift evolutions of $\log_{10}(\dot{N}_{\rm ion}) = (0.08\pm0.01)z + (51.60\pm0.06)$ and $\log_{10}(f_{\rm esc}\xi_{\rm ion}) = (0.07\pm0.01)z + (24.12\pm0.07)$. Late-time values for the ionizing photon production rate density are consistent with both theoretical models and observations. Finally, we measure the evolution of the IGM ionized fraction to find that observed populations of star-forming galaxies are capable of driving reionization in GOODS-S to completion by $z\sim 5.3$ without the need for AGN or other exotic sources, consistent with other studies of the same field. The $20\%$ of UV-brightest galaxies ($M_{\rm UV}<-18.5$) reionize roughly $35\%$ of the survey volume, demonstrating that UV faint LyC emitters are crucial for reionization.
\end{abstract}



\section{Introduction}\label{sec:introduction}

By the end of the Epoch of Reionization, the Universe had undergone its last major phase-transition, and the \ac{IGM} became mostly transparent to the Lyman Continuum (LyC: $\lambda \leq 912$\AA) photons. While current constraints place the mean redshift of reionization at $7.8 \lesssim z \lesssim 8.8$~\citep{Planck:2016}, various observational studies find that this process was complete by a redshift in the range $z\sim5-6$~\citep{Fan:2006,Kulkarni:2019,Becker:2021,Bosman:2022}, contributing to the picture that this process was patchy~\citep{Iliev:2006,Becker:2015, Puchwein:2023}.

Generally, it is believed that the majority of ionizing photons are produced by young, massive stars in galaxies that undergo rapid star formation (e.g.~\citealt{Shapiro:1987, Robertson:2015,Hassan:2018,Rosdahl:18}). However, it is still unclear whether this is driven by a small number of massive sources or from more ``democratic'' contributions from a large number of low-mass galaxies~\citep{Paardekooper:2015,Livermore:2017,Mason:2019,Finkelstein:2019,Naidu:2020,Wu:2024}. Furthermore, certain observational constraints such as the low optical depth to Thompson scattering~\citep{Planck:2016} and high fraction of broad-line \ac{AGN} with large bolometric luminosities among galaxies at redshifts $z \sim 4-6$~\citep{Giallongo:2015,Giallongo:2019} all suggest that the contribution of \ac{AGN} to the ionizing photon budget may be important. However, the late reionization of helium \citep{Kriss:2001,Zheng:2004,Shull:2004,Furlanetto:2008,Shull:2010,Worseck:2016} points to the fact that \ac{AGN} cannot be a dominant component of hydrogen reionization. Furthermore, difficulties in accurately measuring their masses and accretion rates at high redshifts (e.g.~\citealt{Li:2024}) as well as their relative sparsity suggest that they dominate the ionizing photon budget only at lower redshifts $z \lesssim 4$ (e.g.~\citealt{Kulkarni:2019b,Dayal:2020,Trebitsch:2021,Trebitsch:2023}).

Three quantities need to be constrained in order to model the evolution of reionization. First is the UV luminosity function, \rhouv, which describes the number density of sources at a given redshift and UV magnitude. This has been measured from deep imaging surveys (e.g.~\citealt{Bowler:2020,Bouwens:2021,Harikane:2022,Robertson:2023,Varadaraj:2023,Donnan:2023,Donnan:2024}), though the majority of the uncertainty comes from survey completeness (e.g.~\citealt{Robertson:2023}). Similarly, while photometric redshift estimates are occasionally known to be a source of uncertainty\footnote{Particularly in the case of sources at apparently extreme redshifts \citep[e.g.][]{Donnan:2023}.}, these have been found to be generally consistent with spectroscopic confirmations~\citep[e.g.][]{Hainline:2024}.

Second is the ionizing photon production rate per UV luminosity, \xion. This can be predicted either by stellar population synthesis models during \ac{SED} fitting (e.g.~\citealt{Leitherer:1999,Stanway:2018}), or inferred from emission lines such as H$\alpha$ or H$\beta$ (e.g.~\citealt{Maseda:2020,Saxena:2024}) or O~{\small III} equivalent widths~\citep{Chevallard:2018,Tang:2019}. Here, uncertainties are primarily driven by differences in stellar population models (e.g. the presence of binaries, initial mass function, gas geometry, etc.) as well as assumptions about physical conditions in the H~{\small II} regions of sources emitting ionizing photons.

Third, one must account for the fraction of the produced ionizing photons that escape their host galaxy into the IGM (\fesc). Due to the fact that this depends on complex non-linear physics on small scales in the \ac{ISM} (e.g.~\citealt{Kimm:2019,Kimm:2022,Kakiichi:2021}), \fesc{} is highly line-of-sight dependent (e.g.~\citealt{Fletcher:2019,Choustikov:2024b,Yuan:2024} and references therein), and cannot be directly measured at redshifts $z \gtrsim 4$ due to the increasingly neutral \ac{IGM} (e.g.~\citealt{Worseck:2014,Inoue:2014}), 

\fesc\ arguably carries the most uncertainty. The escape fraction of ionizing photons has been studied extensively using both galaxy formation simulations (e.g.~\citealt{Kimm:2014,Xu:2016,Trebitsch:2017,Rosdahl:18,Rosdahl:22,Ma:2020,Saxena:2022a,Giovinazzo:2024}) and observations of low-redshift analogues (e.g.~\citealt{Leitherer:2016,Schaerer:2016,Steidel:2018,Izotov:2018,Izotov:2018b,Flury:2022,Flury:2022b}). In the case of simulations, capturing the production and transfer of LyC photons through a multi-phase \ac{ISM} into a realistic CGM is difficult. To do so requires self-consistently capturing a large dynamical range, along with realistic models for the \ac{ISM} and feedback processes \citep{Kimm:2019,Kimm:2022,Rosdahl:22}. In contrast, it is not clear whether these observed analogues are representative of high-redshift galaxies or plagued by selection effects~(e.g.~\citealt{Katz:2022b,Katz:2023,Brinchmann:2023,Schaerer:2022}). In both cases, the general strategy is to derive indirect diagnostics, that trace physically favourable conditions to LyC production and escape from the \ac{ISM}~\citep{Choustikov:2024}. These include a variety of different indirect tracers, including properties of Ly$\alpha$ emission \citep{Jaskot:2014, Henry:2015,Verhamme:2015,Verhamme:2017,Steidel:2018, Pahl:2021,Naidu:2022,Choustikov:2024b}, high [O~{\small III}]~$\lambda5007$/[O~{\small II}]~$\lambda\lambda3726,3728$ (O$_{32}$) ratios \citep{Nakajima:2014}, particularly negative UV continuum slopes ($\beta$) \citep{Chisholm:2022}, low amounts of UV attenuation \citep{Saldana-Lopez:2022}, Mg~{\small II}~$\lambda\lambda2796,2804$ doublet ratios \citep{Chisholm:2020}, strong C~{\small IV}~$\lambda\lambda1548,1550$ emission \citep{Schaerer:2022,Saxena:2022}, S~{\small II} deficits \citep{Chisholm:2018,Wang:2021}, relative sizes of resonant line surface brightness profiles \citep{Choustikov:2024b,Leclercq:2024} and multivariate models \citep{Mascia:2023, Choustikov:2024,Jaskot:2024a,Jaskot:2024b}.

The primary limitation is that the vast majority of the methods used to infer these properties on a case-by-case basis require spectroscopic information about a given galaxy, which is expensive (particularly in comparison to photometric surveys). Furthermore, studies that are able to make use of photometric observations to constrain certain parameters (primarily by using \ac{SED} fitting) often require making assumptions about the others (particularly \fesc) being constant or evolving on a population level only (e.g.~\citealt{Boyett:2022,Simmonds:2023,Simmonds:2024}). Finally, performing advanced \ac{SED} fitting over a large galaxy sample is a very computationally expensive and time-consuming exercise. 

Given the availability of unprecedented photometric data from \textit{JWST}, the objective of the present work is to develop a model to infer the total escaping output of ionizing photons of a given source based on \textit{JWST} NIRCAM photometric measurements. To do this, we build an \ac{ILI} pipeline developed using \ltuili~\citep{Ho:2024} trained on dust-attenuated mock photometry of a statistical sample of representative high-redshift galaxies from the \sphinx\ simulation~\citep{Rosdahl:22,spdrv1}. This pipeline is able to make accurate and fast predictions for the angle-averaged ionizing photon contribution (\Nion) of individual sources with reliable uncertainties, based on filters used by the \textit{JWST} Advanced Deep Extragalactic Survey\footnote{In principle, the method outlined in this paper is extendable to almost any other \textit{JWST} survey. However, we have focused on \jades\ because it is a particularly deep survey with a large number of filters, making it an ideal proving ground.} (\jades: \citealt{Eisenstein:2023}). Using public data from JADES, we aim to explore the redshift evolution of \Nion\ for a sample of 4,559 photometrically selected galaxies at high redshift. Finally, we will combine these measurements to constrain the evolution of the global ionizing photon production rate (\nion), allowing us to investigate the redshift evolution of reionization in the GOODS-S field.

This paper is arranged as follows. First, in~\Cref{sec:model} we outline \texttt{PHOTONIOn}\footnote{\url{https://github.com/Chousti/photonion.git}}: the pipeline that we have built to predict \Nion\ based on \textit{JWST} photometry. In~\Cref{sec:benchmark} we benchmark our inference pipeline and compare it with another \ac{SED}-fitting method. Next, in~\Cref{sec:results_JADES} we apply this pipeline to a sample of JADES galaxies imaged using \textit{JWST} NIRCam and characterise the ionizing photon contributions of this population of galaxies. Using this, we then compute the evolution of the ionized fraction of the \ac{IGM}. Finally, we discuss caveats of our approach in~\Cref{sec:caveats} before concluding in~\Cref{sec:conclusion}.

Throughout this paper, we assume a flat $\Lambda\mathrm{CDM}$ cosmology with cosmological parameters compatible with \cite{Planck:2014}\footnote{This is chosen to be consistent with the training data from the \sphinx\ simulation \citep{Rosdahl:22}.} as well as a primordial baryonic gas of hydrogen and helium, with mass contents of $X = 0.75$ and $Y = 0.25$, respectively.


\section{\texttt{PHOTONIO\MakeLowercase{n}}: Predicting Escaping Ionizing Luminosity with Implicit Likelihood Inference}\label{sec:model}

\ac{ILI}, also known as \ac{SBI} or \ac{LFI}, is a class of methods to infer the statistical relationship between the observed data ($\boldsymbol{X}$) and the underlying parameters of a model that generated the data ($\boldsymbol{\theta}$). For a thorough review see for example,~\cite{Marin_2011} or~\cite{Cranmer_2020}. To infer $\boldsymbol{\theta}$ from $\boldsymbol{X}$, the Bayes theorem states that the posterior distribution of $\boldsymbol{\theta}$ is given by
\begin{equation}
    \mathcal{P}(\boldsymbol{\theta} | \boldsymbol{X}, I)
    \propto
    \mathcal{L}(\boldsymbol{X} | \boldsymbol{\theta}, I) \pi(\boldsymbol{\theta} | I),
\end{equation}
where $\mathcal{L}(\boldsymbol{X} | \boldsymbol{\theta}, I)$ is the likelihood of the data given the model, $\pi(\boldsymbol{\theta} | I)$ is the prior distribution of the model parameters, and $I$ denotes the remaining information required to specify the model. In many applications, the likelihood function may be unknown or computationally intractable while the mapping $\boldsymbol{\theta} \rightarrow \boldsymbol{X}$ is available. Thus, \ac{ILI} relies on a ``simulator'' which can or has generated such synthetic data to populate a high-dimensional space of model parameters and observed data (see Figure 1 in~\citealt{Ho:2024}). In turn, this can be used to infer the \textit{distribution} of plausible model parameters that may have generated the observed data by slicing the space at the observed data.

In this work, we opt for the neural posterior estimation method~\citep{Papamakarios_2016, Greenberg_2019}, which directly emulates the posterior distribution. This is particularly suitable because in our case we have a single model parameter (\Nion) and a $13$-dimensional space of observed data. Specifically, to be consistent across all sources, we use photometric magnitudes in the F115W, F150W, F200W, F277W, F335M, F356W, F410M and F444W filters normalised by the apparent UV magnitude ($m_{\rm AB}^{1500}$), three colours (F115W-F150W, F150W-F277W, and F277W-F444W) as well as $m_{\rm AB}^{1500}$ and redshift. However, other flavours of \ac{ILI} exist such as the neural likelihood estimation~\citep{Alsing_2018, Papamakarios_2018} or the neural ratio estimation~\citep{Hermans_2019}.

In case of neural posterior estimation, we wish to approximate the ``true'' posterior $\mathcal{P}(\boldsymbol{\theta} | \boldsymbol{X}, I)$ with the neural posterior $\hat{\mathcal{P}}(\boldsymbol{\theta} | \boldsymbol{X}, I)$ while only having access to samples $\mathcal{D}_{\rm train} = \{\boldsymbol{X}_i, \boldsymbol{\theta}_i\} $ from the simulator. The neural posterior may be decomposed as
\begin{equation}
    \hat{\mathcal{P}}(\boldsymbol{\theta} | \boldsymbol{X}, I)
    =
    \frac{\pi(\boldsymbol{\theta} | I)}{p(\boldsymbol{\theta} | I)} q_{\boldsymbol{w}}(\boldsymbol{\boldsymbol{\theta}} | \boldsymbol{X}, I),
\end{equation}
where $\pi (\boldsymbol{\theta} | I)$ is the prior distribution of $\boldsymbol{\theta}$, $p(\boldsymbol{\theta} | I)$ is the proposal prior representative of the distribution of $\boldsymbol{\theta}$ in the simulated (training) data (which down-weights over-represented values of $\boldsymbol{\theta}$), and $q_{\boldsymbol{w}}(\boldsymbol{\boldsymbol{\theta}} | \boldsymbol{X}, I)$ is the neural network output. Although typically $q_{\boldsymbol{w}}$ is modelled with a normalizing flow~\citep{Papamakarios_2019}, in our case $\boldsymbol{\theta}$ is only 1-dimensional and thus we opt for a mixture density network~\citep{Bishop_1994}. Specifically, we use a Gaussian mixture density network to model $q_{\boldsymbol{w}}$, where the neural network with weights and biases $\boldsymbol{w}$ outputs the parameters of the mixture (mean and standard deviation of each component of the mixture). Furthermore, we also assume the prior and proposal distributions to be identical. During training, the network parameters $\boldsymbol{w}$ are optimized using a loss function
\begin{equation}\label{eq:loss}
    L = - \sum_{i \in \mathcal{D}_{\rm train}} \log \hat{\mathcal{P}}(\boldsymbol{\theta}_i | \boldsymbol{X}_i, I),
\end{equation}
introduced by~\cite{Papamakarios_2016}. We implement the neural posterior estimator using \ltuili\footnote{\url{https://github.com/maho3/ltu-ili}}\ pipeline introduced by~\cite{Ho:2024}.

\begin{figure*}
    \includegraphics{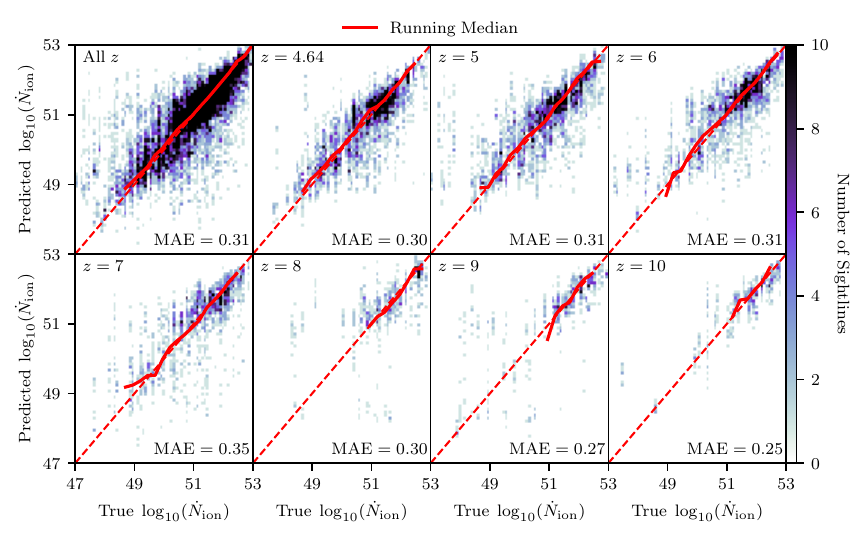}
    \caption{Histogram of \protect\Nion\ predicted using the \ac{ILI} pipeline applied to mock dust-attenuated photometry of \protect\sphinx\ galaxies as well as the true values computed using \protect\texttt{RASCAS}, broken down by redshift bins. We include the running medians in red as well as the median absolute error (MAE) for each redshift bin, showing how well the model performs in this validation experiment.}
    \label{fig:model}
\end{figure*}

In order to train the model, we use 13,800 mock line-of-sight dust-attenuated photometric observations of star-forming galaxies from \sphinx\ \citep{Rosdahl:18,Rosdahl:22}, a cosmological radiation hydrodynamical simulation of reionization in a $20~\rm{cMpc}$ box with sufficient resolution to resolve the multi-phase \ac{ISM} in a large population of constituent galaxies. Specifically, this data-set consists of a sample of $1,380$ star-forming galaxies at $z = 10, 9, 8, 7, 6, 5, {\rm \;and\;}4.64$. These galaxies were selected to have $10~\mathrm{Myr}$ averaged ${\rm SFR} \geq 0.3~\mathrm{M}_\odot{\rm yr}^{-1}$, so that they form a representative sample of galaxies that could be observed by a flux-limited \textit{JWST} survey~\citep{Choustikov:2024}. Available as part of the \sphinx\ Public Data Release (SPDRv1,~\citealt{spdrv1}), each galaxy has been post-processed with \texttt{RASCAS} \citep{Michel-Dansac:2020} to simulate the self-consistent generation and propagation of an \ac{SED} consisting of the stellar continuum, nebular continuum and nebular emission lines. A peeling algorithm \citep[e.g.][]{Yusuf_Zadeh_1984,Zheng_2002,Dijkstra_2017} was used to mock observe these dust-attenuated \acp{SED} along ten consistent lines-of-sight, producing photometric images and magnitudes in \textit{JWST} NIRCam filters. Comparisons between mock \sphinx\ and \jades\ photometry and colour have been carried out, confirming that this is a representative sample (see Figs. 15 and 16 of ~\citealt{spdrv1}). A complete description of the methods used to generate this data-set are provided in \cite{spdrv1} and \cite{Choustikov:2024}.

We train the model to predict $\log_{10} \dot{N}_{\rm ion}$, apply standard scaling to both the features and targets, and opt for a $20$-$80\%$ test-train split by galaxies, not by individual lines-of-sight, to ensure that a single galaxy is not present in both splits. Furthermore, to make training more robust, we only use galaxies with $f_{\rm esc}\geq10^{-6}$ to remove a small tail of outliers\footnote{Doing so improves the general performance of the model, as machine learning methods can struggle to reproduce outliers.} and ensure that the full distribution of \Nion\ values are represented in the training set. We use \texttt{Optuna}~\citep{optuna} to optimize the following hyperparameters: number of hidden features in the network, number of mixture components, optimizer learning rate, training batch size and the early stopping criterion. We run \texttt{Optuna} for $1,000$ trials to find the best hyperparameters and optimize the mean of~\cref{eq:loss} in a $10$-fold cross-validation across galaxies. We list the selected hyperparameters to predict $\log \dot{N}_{\rm ion}$ from the \jades\ filters and redshift in~\cref{table:hyperparameters}.

\begin{table}
    \centering
    \begin{tabular}{l|c|l}
        Hyperparameter                  &  Optimized & Value   \\ \hline \hline
        Number of hidden features       &  \cmark    & 21 \\
        Number of mixture components    &  \cmark    & 3 \\ \hline
        Optimizer learning rate         &  \cmark    & $\num{8.932e-4}$ \\
        Training batch size             &  \cmark    & 45 \\
        Early stopping criterion        &  \cmark    & 13 \\
        Validation fraction             &  \xmark    & $0.2$   \\
        Gradient norm clipping          &  \xmark    & $5$     \\
        \hline
    \end{tabular}
    \caption{Selected hyperparameters of the \ac{ILI} model predicting $\log \dot{N}_{\rm ion}$ from \jades\ filters and source redshift. The hyperparameter naming follows the \ltuili\ interface and we outline the hyperparameter optimization routine in~\cref{sec:model}.}
    \label{table:hyperparameters}
\end{table}

Having trained the model, we can draw samples from~$\hat{\mathcal{P}}(\boldsymbol{\theta} | \boldsymbol{X}, I)$. When testing the model on simulated data without uncertainties, we either draw $1,000$ samples from the learnt posterior or summarize those draws with the maximum posterior value and an asymmetric $1\sigma$ uncertainty around it. On the other hand, when applying the model to observational data with uncertainties, we assume the uncertainties to be Gaussian such that $\boldsymbol{X} \pm \Delta \boldsymbol{X}$ and re-sample $\boldsymbol{X}$ $500$ times, each time sampling $1,000$ draws from the posterior. In doing so, we propagate both the model and photometric uncertainties into the prediction of $\boldsymbol{\theta}$.

In~\cref{fig:model} we compare the predicted \Nion\ with the true \Nion\ of \sphinx\ galaxies, isolating the sample of mock observations at each redshift in our sample. We note that we train a single model with redshift as a feature as opposed to training a separate model for each redshift bin. In all cases, we find that the running median of the distribution matches the one-to-one line well, with the complete sample having a \ac{MAE} of $0.31~\mathrm{dex}$. The model performs particularly well for sources with $\log_{10}(\dot{N}_{\rm ion} / [{\rm photon}/{\rm s}]) > 51$, struggling more with the LyC-dimmest sources at the highest redshifts for which training data is limited. For completeness, we perform a variety of other benchmark tests on the model. These are discussed in~\cref{sec:validate}.

Finally, we highlight that this method allows us to predict the \textit{global} escaped ionizing luminosity of high-redshift galaxies without having to dust-correct observations or assume some model for the LyC escape fraction. As a result, this method is completely self-consistent, simple and efficient; as compared to traditional \ac{SED}-fitting methods.


\section{Benchmarking the Model}\label{sec:benchmark}

\subsection{Comparison with a Standard SED-Fitting Method}\label{sec:bagpipes}

\begin{figure*}
    \includegraphics{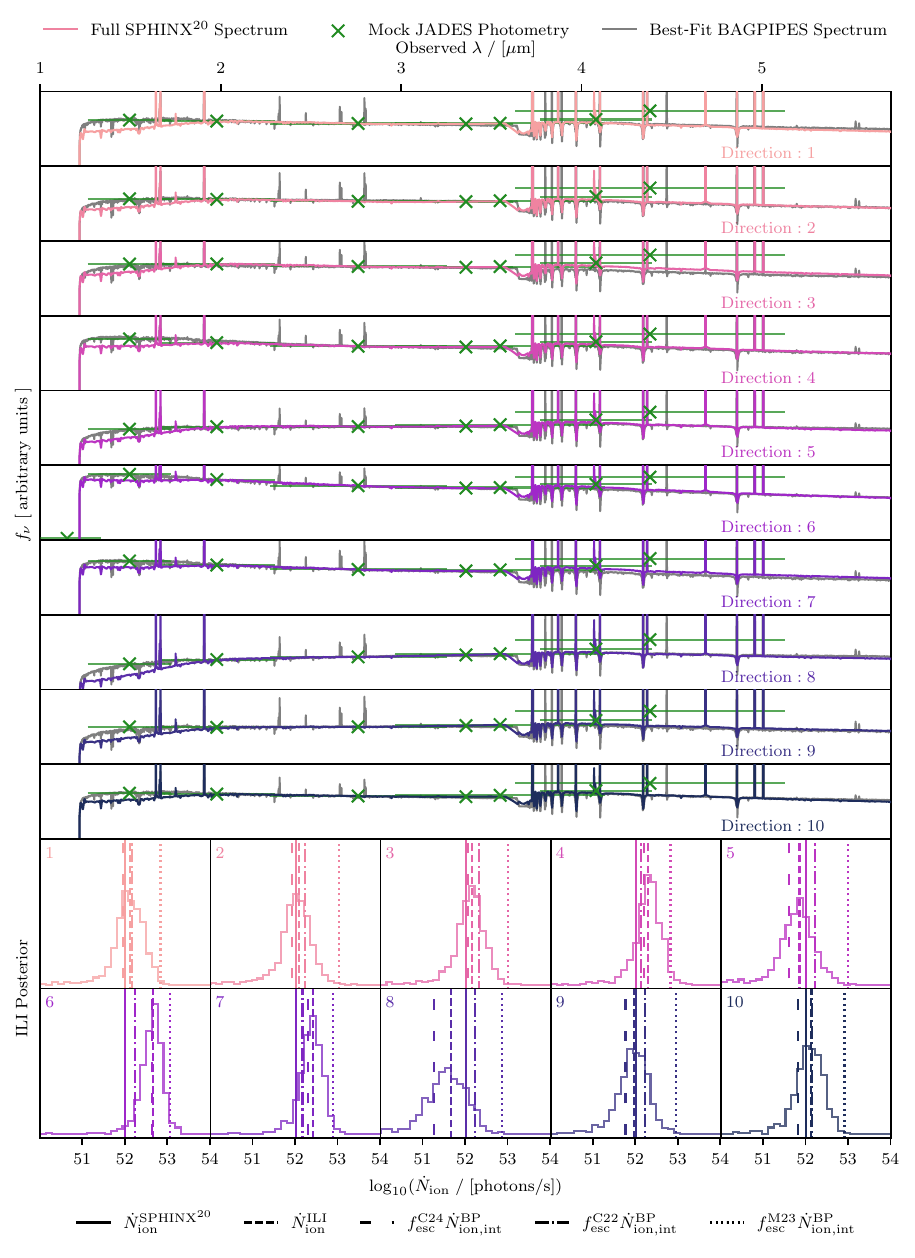}
    \caption{Comparison between the implicit likelihood inference (ILI) and \texttt{BAGPIPES} methods to inferring \Nion\ for a random test-set \sphinx\ galaxy at redshift $z = 9$. In the \textit{long} panels, we include the mock \ac{SED} (colour), mock \textit{JWST} NIRCam photometry in the \jades\ filters (green), and best-fit \texttt{BAGPIPES} \ac{SED} (gray) for each line of sight. On the \textit{bottom}, we include the full posterior distributions for \Nion\ as sampled by the \ac{ILI} pipeline, along with the true (\textit{solid}) value, best-fit \ac{ILI} (\textit{dashed}) and best-fit \texttt{BAGPIPES} predictions (\textit{loosely dashed, dot-dashed, dotted}) using \fesc\ models from \protect\cite{Choustikov:2024,Chisholm:2022,Mascia:2023} respectively. We note that $\dot{N}_{\rm ion,int}^{\rm BP} = \dot{N}_{\rm ion,int}^{\rm BAGPIPES}$ as a shorthand.}
    \label{fig:hero}
\end{figure*}

\begin{figure*}
    \includegraphics{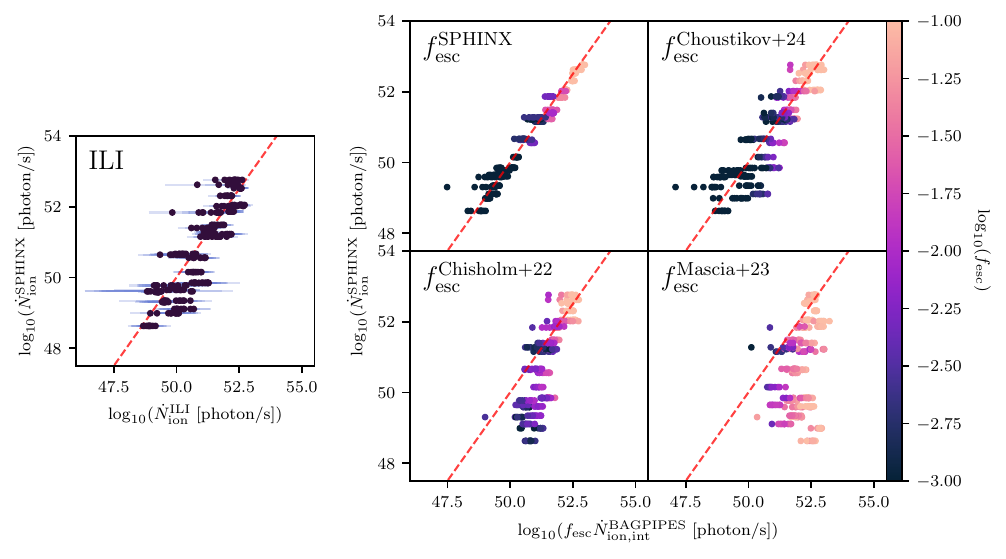}
    \caption{(\textit{Left}): Comparison between the true value of escaped ionizing luminosity ($\dot{N}_{\rm ion}^{\rm SPHINX^{20}}$) of all \sphinx\ galaxies computed using \texttt{RASCAS} with predictions from the \ac{ILI} pipeline. We include asymmetric $1\sigma$ error bars based on the \ac{ILI} posteriors, as discussed in~\protect\Cref{sec:model}. \textit{Right}: Comparison between the true value of \Nion\ with predictions using the best-fit \texttt{BAGPIPES} \ac{SED}. Here, we use LyC escape fractions computed using methods proposed by~\protect\cite{Choustikov:2024} (\textit{top right}), \protect\cite{Chisholm:2022} (\textit{bottom left}),and~\protect\cite{Mascia:2023} (\textit{bottom right}), as well as the true values computed by \texttt{RASCAS} (\textit{top left}). Everywhere, the one-to-one relation is shown in red. In each case, points are coloured by the value of the escape fraction used. This  highlights that while \texttt{BAGPIPES} has recovered the intrinsic ionizing luminosity well, much of the uncertainty in measurements of \Nion\ is dominated by the escape fraction prescription, even if one assumes access to spectroscopic data. In contrast, the ILI method is able to account for this with only photometric data.}
    \label{fig:bagpipes2}
\end{figure*}

While we have shown that our model is accurate with well-behaved uncertainties, it is important to compare the efficacy of this method to that of a traditional \ac{SED}-fitting code. To this end, we choose to use the default version of \texttt{BAGPIPES}~\citep{bagpipes} owing to its ease of deployment. For a random sample of $30$ galaxies from the \sphinx\ database we use \texttt{BAGPIPES} to find the best-fit model spectrum for each set of line-of-sight mock photometry. We note that to ensure that this is a fair test, we only sample galaxies from the test set of the \ac{ILI} model introduced in~\Cref{sec:model}. For our Stellar Population Synthesis models, we use the 2016 version of the BC03 templates \citep{Bruzual:2003, Chevallard:2016} with a \cite{Kroupa_2001} IMF. Nebular (both line and continuum) emission is accounted for with \texttt{CLOUDY} \citep{Ferland:2017}. Our star formation history (SFH) is taken to be non-parametric, following the continuity prior introduced by \cite{Leja:2019} with time bins set to $[0., 10., 25., 50., 100., 250., 500., t_{z}]\;{\rm Myr}$ where $t_z$ is the age of the universe at redshift $z$. Following \cite{Tacchella:2022}, we allow the SFH to be more bursty by adjusting parameters in the continuity prior Student's t-distribution to $\sigma=1$ and $\nu=2$. This SFH model was recently shown to recover the stellar masses of \sphinx\ galaxies well \citep{Cochrane:2024}. For the total stellar mass and metallicity of each galaxy, we use uniform priors of $\log_{10}(M/M_{\odot}) \in [0, 10]$ and $\log_{10}(Z/Z_{\odot}) \in [-3, 1]$ respectively. Accounting for dust, we fit an SMC dust law \cite{Gordon:2003} with uniform V-band attenuation priors of $A_V \in [0,2]$. In line with other work \citep[e.g.][]{Cochrane:2024}, we also assume that birth clouds attenuate young stars ($t_{\rm BC}<10{\rm Myr}$) twice as much as older stellar populations that have had a chance to clear their local ISMs \citep{Calzetti:1994}. For simplicity, we assume an ionization parameter of $\log_{10}(U) = -2$. Finally, we take the `best-case-scenario' approach \citep[e.g.][]{Narayanan:2024} by fixing the redshift at the true value from the simulation.

In order to compute the ionizing photon flux of each galaxy, $\dot{N}_{\rm ion,int}$, we first start by converting the intrinsic stellar SED of each galaxy (accounting for both the birth cloud and older stars) into a photon flux. We then integrate these over the rest-frame wavelengths of $505\AA\leq\lambda< 912$, with a lower bound set by the first ionization energy of helium. To model the escape fraction, we again take the `best-case-scenario' approach and assume that we have spectroscopy of each source along each line of sight, allowing us to use the spectroscopic properties typically needed to infer the LyC escape fraction. we use increasingly advanced models based on the UV spectral index ($\beta$) \citep[$f_{\rm esc}^{\rm C22}$,][]{Chisholm:2022}, the effective half-light radius observed in the F115W filter ($R_e$), $\beta$, and the ratio of [O~{\small III}]~$\lambda$5007/[O~{\small II}]~$\lambda\lambda$3726,3728 (O$_{32}$) \citep[$f_{\rm esc}^{\rm M23}$,][]{Mascia:2023}. Finally, we use the generalized linear model given by Equation 4 of \cite{Choustikov:2024} which takes $\beta$, the UV-attenuation (E(B-V)), H$\beta$ flux, absolute UV magnitude (M$_{\rm UV}$), the ratio of ([O~{\small III}]~$\lambda$5007+[O~{\small II}]~$\lambda\lambda$3726,3728)/H$\beta$ ($\rm R_{23}$), and O$_{32}$ to provide $f_{\rm esc}^{\rm C24}$. All of the necessary data is provided by the \sphinx\ Public Data Release v1 \citep{spdrv1}. In each case, all line fluxes have been dust-corrected using the SMC dust law \cite{Gordon:2003} and we neglect contributions from the nebular continuum which can redden observed UV galaxy slopes \citep{Katz:2024}. The product of the calculated $\dot{N}_{\rm ion,int}$ and one of the inferred escape fractions is taken as the final prediction. We note that there is a small inconsistency in this approach, due to the fact that \texttt{BAGPIPES} implicitly assumes that the escape fraction of `birth cloud' stars is zero during the fitting stage. We leave exploration of this effect to future work, but suggest that other SED-fitting codes have loosened this constraint \citep[e.g. PROSPECTOR,][]{prospector}. Finally, we repeat this process by also predicting \Nion\ using \texttt{PHOTONIOn} as described in~\Cref{sec:model}.

This entire process is demonstrated in~\cref{fig:hero}, where we display all of the necessary information for all $10$ lines-of-sight for a randomly selected test-set \sphinx\ galaxy at redshift $z = 9$. On the \textit{top}, we show the full mock \sphinx\ \ac{SED} (\textit{in colour}), the mock \textit{JWST} NIRCam photometry in the \jades\ filters (\textit{green}), as well as the best-fit \texttt{BAGPIPES} \ac{SED} (\textit{gray}), confirming that it is a good match. On the \textit{bottom}, we show the full \ac{ILI} posterior distribution for each sight line (with matched colours). In each case, we include the true value of \Nion, computed directly from \texttt{RASCAS} ($\dot{N}_{\rm ion}^{\rm SPHINX^{20}}$, \textit{solid}), the best-fit \ac{ILI} prediction ($\dot{N}_{\rm ion}^{\rm ILI}$, \textit{dashed}), along with the best-fit \texttt{BAGPIPES} predictions ($\dot{N}_{\rm ion,int}^{\rm \texttt{BAGPIPES}}$) with $f_{\rm esc}^{\rm C24}$ (\textit{loosely dashed}), $f_{\rm esc}^{\rm C22}$ (\textit{dot-dashed}), and $f_{\rm esc}^{\rm M23}$ (\textit{dotted}). Here, we find that the \ac{ILI}-inferred values are typically much more accurate and consistent than those inferred from \texttt{BAGPIPES} model \acp{SED}, despite the fact that \texttt{BAGPIPES} is inferring the \ac{SED} well, as we discuss further below. Furthermore, it is clear to see that the lines-of-sight for which this is not the case (6 and 8) are significantly bluer and dustier than the others, respectively. In these cases, the \ac{ILI} pipeline performs as expected and produces more uncertain, broader posteriors that tend to have skewed predictions for \Nion, with bluer(redder) sightlines over(under)-predicting \Nion. However, the larger error bars confirm that the model is behaving as required. Finally, it is interesting to compare the relative success of the various \texttt{BAGPIPES} results. Here, we find that these methods tend to have dramatically different estimates for \Nion\ for a given source, due to disagreements in the inferred escape fractions. In the case of this example, the \cite{Choustikov:2024} and \cite{Chisholm:2022} methods both perform well, while that of \cite{Mascia:2023} struggles and tends to systematically over-predict \Nion. We will continue to discuss these differences below.

Now, we proceed to test how well these approaches can recover the escaped ionizing luminosities of \sphinx\ galaxies. Figure \ref{fig:bagpipes2} shows this in full. On the \textit{left}, we present a comparison between the true values of $\dot{N}_{\rm ion}^{\rm SPHINX^{20}}$ compared to those predicted by \texttt{PHOTONIOn}, along with the associated uncertainties. In contrast on the \textit{right}, we show the same for the values of \Nion\ inferred using \texttt{BAGPIPES} with a variety of escape fraction prescriptions. These include the true value from the simulation (\textit{top left}), the \cite{Choustikov:2024} approach (\textit{top right}), the \cite{Chisholm:2022} method (\textit{bottom left}), and the prescription of \cite{Mascia:2023} (\textit{bottom right}). In each case, we show the one-to-one relation in red and color points by the value of the escape fraction applied. In doing so, we reiterate that while the ILI method implicitly accounts for the escape fraction using only photometric data, these approaches assume a `best-case-scenario' where we also have access to spectroscopic data. Finally, horizontal streaks of points show uncertainty picked up in observing these galaxies from multiple sight-lines.

We find that the ILI approach is able to recover the one-to-one relation well, with the same degree of scatter shown in Figure \ref{fig:model}. Next, as a sanity check, we find that when the true escape fraction is used, \texttt{BAGPIPES} is able to recover the intrinsic ionizing luminosities of these galaxies very well. However, it is clear that as soon as some prescription for the escape fraction is applied, this introduces a significant amount of scatter in the predicted value of \Nion. We find that while the \cite{Choustikov:2024} approach is at least able to follow the one-to-one line across all values of \Nion\ studied (though we caution that this method was fitted to \sphinx\ galaxies), the approaches suggested by \cite{Chisholm:2022} and especially \cite{Mascia:2023} break down. Particularly, both of these approaches can severely over-predict \Nion\ for the weakest LyC leakers, by as much as 4 dex in the worst case. Interestingly, while the $\beta$-slope method of \cite{Chisholm:2022} is at least able to perform well for galaxies with $\dot{N}_{\rm ion}\gtrsim10^{51}{s^{-1}}$, the introduction of dependencies on O$_{32}$ and $R_e$ appears to make even this difficult in the latter model. We believe this is primarily due to the positive correlation with O$_{32}$, which has been shown to break down for the strongest leakers \citep{Choustikov:2024}, while the negative correlation with $R_e$ has been shown to hold true \citep[see Figure 14 of][]{Choustikov:2024b}.

Nevertheless, it is important to reiterate that while the ILI approach produces a similar amount of scatter to that of \texttt{BAGPIPES} with the \cite{Choustikov:2024} prescription for \fesc, it is able to do so with no access to spectroscopic data whatsoever. This is crucial, as it becomes necessary to analyze large photometric samples, owing to their completeness.

\begin{figure}
    \includegraphics{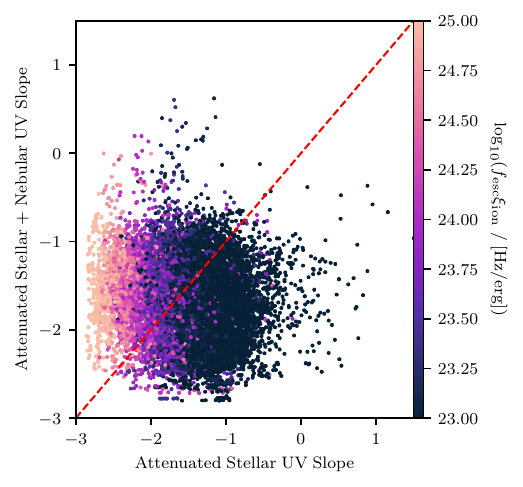}
    \caption{Dust-attenuated UV continuum slopes for \sphinx\ galaxies with and without accounting for the nebular continuum contribution, coloured by $f_{\rm esc}\xi_{\rm ion}$. $\beta$ can be a reliable indicator for the escape fraction of galaxies only if you can disentangle the stellar and nebular contributions to the UV continuum.}
    \label{fig:UV_slopes}
\end{figure}

Interestingly, in the process of this work it was discovered that $\beta$ slope-based methods of inferring the LyC escape fraction perform worse in galaxies with non-negligible nebular continuum contributions. To explore this further, in Figure \ref{fig:UV_slopes} we show the dust-attenuated UV continuum slopes of all \sphinx\ galaxies computed with and without the nebular continuum, coloured by the escaping ionizing production efficiency, $f_{\rm esc}\xi_{\rm ion}$. Several things are immediately clear. First, galaxies with UV slopes significantly reddened by the nebular continuum (i.e. in the \textit{top left} of the figure) are releasing the largest number of ionizing photons per stellar UV luminosity. Secondly, we find that while the UV slopes of the stellar-only continuum are indeed good predictors of LyC escape \citep[][]{Chisholm:2022,Choustikov:2024}, UV slopes measured on the full (stellar and nebular) continuum perform worse, with the strongest effective LyC producers appearing with UV slopes in the range of $-2.5<\beta<-0.5$, including red systems which might traditionally be ignored \citep{Saxena:2024b}. Therefore, it is clear that if this approach is to be used, it is important to first disentangle the stellar continuum from the total observed continuum emission. Finally, it is important to note that the nebular continuum can impact measurements of both $\xi_{\rm ion}$ as well as $M_{\rm UV}$. We refer the interested reader to a complete discussion in \cite{Katz:2024}. While the impact of nebular LyC emission has begun to be explored \citep{Simmonds:2024b}, we leave studies of the impact of nebular continuum light on LyC diagnostics to future work.

\begin{figure*}
    \includegraphics[width=\textwidth]{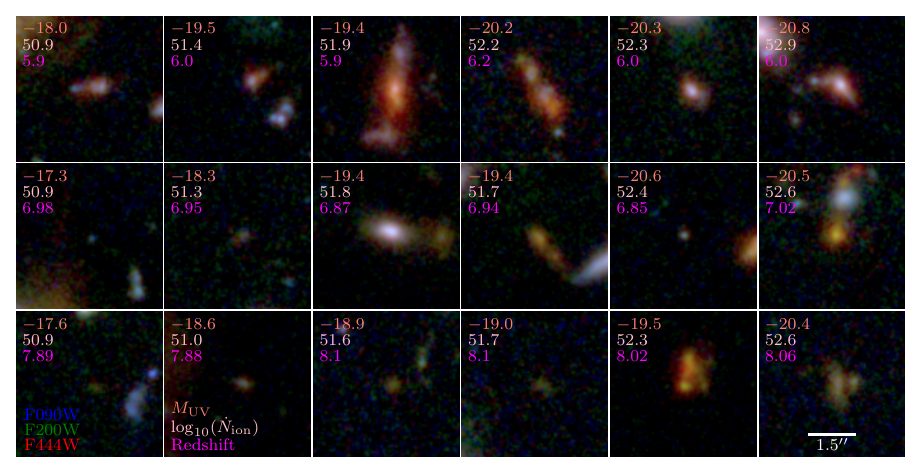}
    \caption{Thumbnail images of 18 galaxies in GOODS-S imaged by \textit{JWST} NIRCam as part of \jades\ \protect\citep{Eisenstein:2023}. RGB images are made using F444W in the red channel, F200W in the green and F090W in the blue channel. For each galaxy, we provide the absolute UV luminosity, \Nion\ predicted by our ILI pipeline as well as photometric redshift. Galaxies are shown in order of their ionizing photon contributions at each redshift.}
    \label{fig:post_stamps}
\end{figure*}

\subsection{All Roads Lead Back to the Escape Fraction}
\label{sec:fesc}

As is perhaps unsurprising, it is clear that the largest source of uncertainty in any such work is in the prescription used to model the LyC escape fraction. This agrees with a large body of work which has pointed to the fact that there is significant scatter in any correlation between observables and \fesc\ \citep[e.g.][]{Flury:2022b,Yeh:2023,Giovinazzo:2024,Yuan:2024,Choustikov:2024,Choustikov:2024b}.

Much of this comes from the fact that predicting \fesc\ requires us to infer small-scale ISM conditions based on large-scale aggregate galaxy properties, which are not always representative of LyC-leaking regions \citep[e.g.][]{Solhaug:2024}. Furthermore, if the goal is to understand a galaxy's contribution to reionization, it is crucial to model the \textit{angle-averaged} escape fraction as opposed to the \textit{line-of-sight} escape fraction \cite[see Appendix B of][]{Choustikov:2024b}. This in itself can introduce biases, given the fact that many of the tracers used are themselves highly line-of-sight dependent \citep[e.g. Ly$\alpha$ properties,][]{Blaizot:2023}. Furthermore, all models trained on line-of-sight data \citep[e.g.][]{Chisholm:2022,Mascia:2023,Jaskot:2024a,Jaskot:2024b} will have this bias built-in, tending to regress to the mean as is seen in Figure \ref{fig:bagpipes2}. Finally, it is not sufficient to discuss the production and escape of ionizing photons as two separate quantities, owing to the fact that both quantities depend on the age of the stellar population producing them and are therefore non-trivially correlated \citep[e.g.][]{Menon:2024}. Indeed, doing so and applying population-averaged values can lead to over-predictions in the estimated ionizing luminosity of galaxies (see Figure \ref{fig:nion_z}).

While the impact of these uncertainties for traditional SED-fitting methods were all shown clearly in Figure \ref{fig:bagpipes2}, it is important to reiterate that for this analysis we assumed that we also had access to spectroscopic data for each galaxy and could reliably distinguish contributions from the nebular continuum. As a result, while \texttt{BAGPIPES} was able to accurately recover the intrinsic ionizing luminosity of these galaxies, it was let down by the unreliability of standard \fesc\ prediction methods. In reality, even if such an approach worked, doing an exercise like this for a large number of observations would be prohibitively expensive as only photometric data is likely to be available for the vast majority of dim high-redshift sources. This further highlights the necessity for the approach presented here with \texttt{PHOTONIOn}, which is designed to accomplish this based on photometry alone.

Finally, it is particularly interesting that the majority of standard approaches tested \citep{Chisholm:2022,Mascia:2023} tended to over-predict the ionizing luminosities of the LyC-dimmest galaxies. This may be significant for discussions of whether reionization is driven by the brightest or faintest leakers (e.g.~\citealt{Finkelstein:2019, Naidu:2020,Simmonds:2024}), by artificially boosting the impact of the weakest sources. For example, \cite{Munoz:2024} recently discussed the possibility that present-day \textit{JWST} observational constraints on $\rho_{\rm UV}$ \citep[e.g.][]{Donnan:2024} and \xion\ \citep[e.g.][]{Simmonds:2024} suggest that there might be \textit{too} many LyC photons, reionizing the Universe too early for alternative probes of the \ac{IGM}. To do this, \cite{Munoz:2024} extrapolated the \fesc\ model of \cite{Chisholm:2022}. However, the results of~\cref{fig:bagpipes2} suggest that doing so is likely to significantly over-estimate the number of escaping LyC photons. This effect will certainly contribute to this conclusion by artificially boosting the ionizing contribution of high-redshift galaxies.

In the future, it is crucial that better care be taken in discussing the estimated escape fractions of high-redshift galaxies. In principle, it is important to fold in as much information as is known about a given galaxy in order to jointly understand both the production and angle-averaged escape of ionizing photons. The method presented in this work represents an attempt at doing just this, for the first time presenting a tool to self-consistently infer the production and escape of ionizing photons from high-$z$ galaxies based on photometry alone, designed to be unencumbered by many of the issues raised above. Now that we have \texttt{PHOTONIOn}, we proceed to apply it to real data to study the evolution of reionization within a deep photometric survey.


\section{Predicting the Escaped Ionizing Luminosities for a Population of \jades\ 
Galaxies}\label{sec:results_JADES}

\subsection{Application to \jades\ NIRCam Data}\label{sec:jades}

\begin{figure}
    \includegraphics{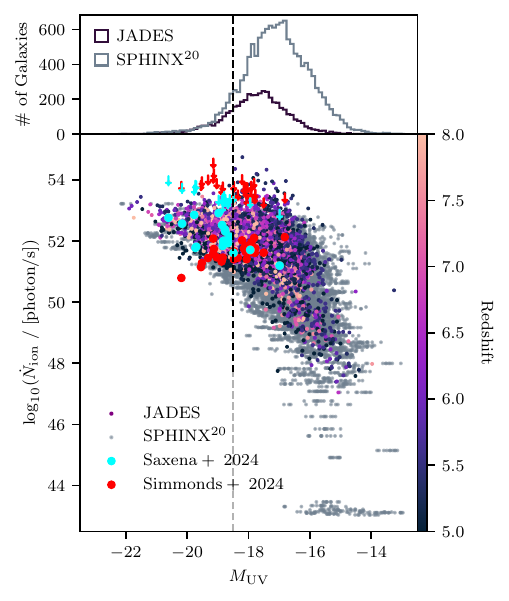}
    \caption{Escaping ionizing luminosity as a function of observed absolute UV magnitude for our sample of \jades\ (\textit{coloured by redshift}) and \sphinx\ (\textit{gray}) galaxies. Other observational data from \protect\cite{Saxena:2024} and \protect\cite{Simmonds:2024} are included for comparison in \textit{cyan} and \textit{red}. In each case, we follow each papers' method to predict \fesc. For the former, we use the multivariate model from \protect\cite{Choustikov:2024}, while in the latter we infer \fesc\ from the absolute UV magnitude based on the relation from \protect\cite{Anderson:2017}. We also include a histogram of the observed absolute UV magnitudes for our sample of \jades\ (\textit{black}) compared to \sphinx\ (\textit{gray}) galaxies. Finally, the cut of UV-bright and UV-dim galaxies ($M_{\rm UV} = -18.5$) used elsewhere in this paper is also shown as a dashed line. Galaxies brighter than this value account for $\sim20\%$ of the sample.}
    \label{fig:muv_nion}
\end{figure}

\begin{figure*}
    \includegraphics{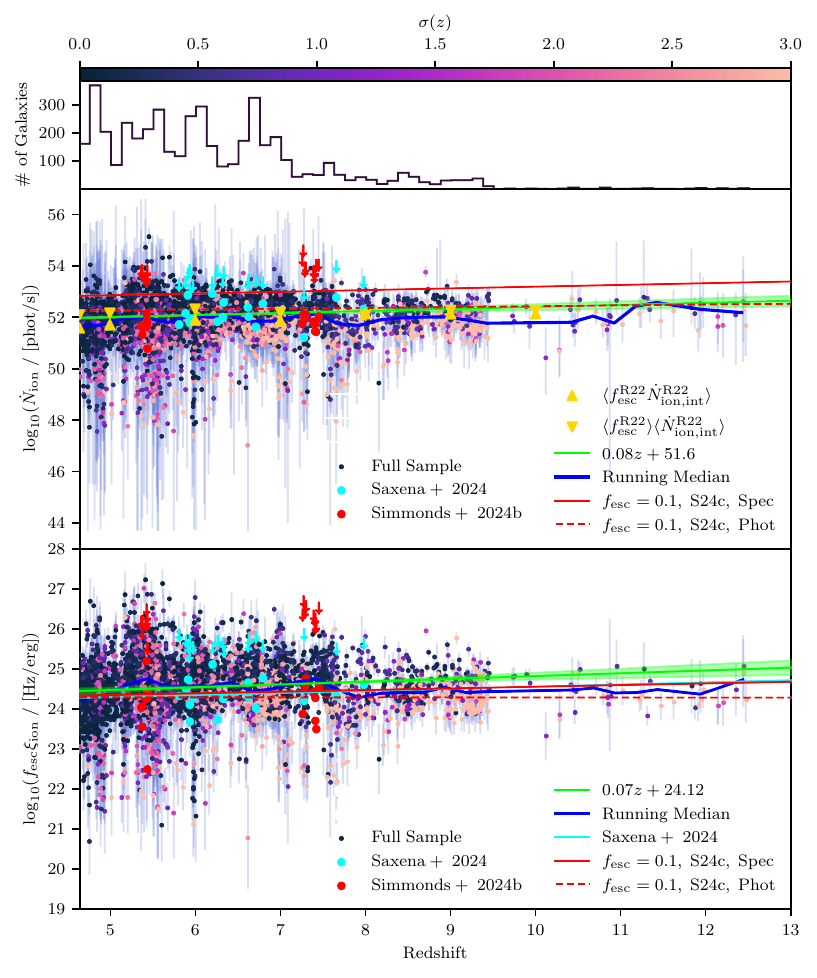}
    \caption{\textit{Middle}: Predicted escaping ionizing luminosities of \textit{JWST} galaxies coloured by their photometric redshift uncertainties. Error bars are produced by resampling the model and photometric uncertainties, as described in the text. We include a running median (\textit{blue}) as well as a line of best-fit (\textit{lime}), computed using \texttt{ROXY} \protect\citep{roxy}. For comparison, we include data from \protect\cite{Saxena:2024}, \protect\cite{Simmonds:2024}, as described in~ \protect\cref{fig:muv_nion}, as well as lines of best-fit derived in \protect\cite{Simmonds:2024c} (\textit{red}) for their spectroscopic (\textit{solid}) and photometric (\textit{dashed}) samples with an assumed escape fraction of $10\%$. Finally, we also include global averages for escaped \Nion, computed at each redshift in \sphinx\ (\textit{gold}). This is to demonstrate the over-prediction which comes from studying these two quantities in isolation.
    \textit{Bottom}: As above but for the escaped ionizing production efficiency, $f_{\rm esc}\xi_{\rm ion}$. Again, we include lines of best fit, running median, observational data \protect\citep{Saxena:2024,Simmonds:2024}, and observational fits \protect\citep{Saxena:2024,Simmonds:2024c}. In both cases, we find that the number of ionizing photons produced and released into the IGM increases gently with redshift, in accordance with observational data.}
    \label{fig:nion_z}
\end{figure*}

\begin{figure*}
    \includegraphics{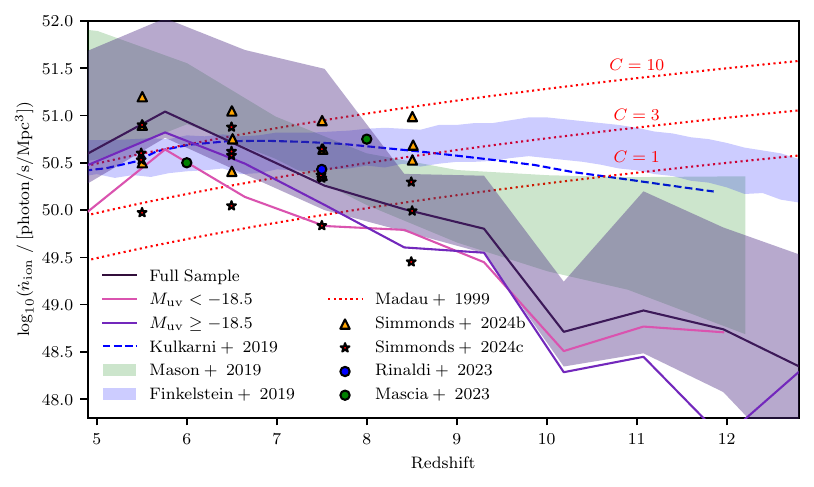}
    \caption{Number density of ionising photons produced and emitted into the IGM within the GOODS-S field as a function of redshift, based on ILI predictions for individual galaxies. We include lines for the entire sample (\textit{black}) as well as for UV-bright galaxies ($M_{\rm UV} < -18.5$; \textit{magenta}) and UV-dim galaxies ($M_{\rm UV} \geq -18.5$; \textit{purple}). For the full sample, we also include uncertainties computed by resampling both photometric and model uncertainties (\textit{dark purple shaded region}). Comparisons with a Bayesian-inferred history \protect\citep{Mason:2019}, theoretical models \protect\citep{Kulkarni:2019,Finkelstein:2019}, and observational data \protect\citep{Rinaldi:2023, Mascia:2023b,Simmonds:2024,Simmonds:2024c} are also provided, as well as an analytical estimate of the number of photons required to ionize the neutral IGM for various clumping factors $C \in \{1, 3, 10\}$ \protect\citep{Madau:1999}.}
    \label{fig:nion_goodss}
\end{figure*}

We now apply our \ac{ILI} pipeline to real data to infer the ionizing photon luminosity of photometrically-observed galaxies. To do so, we use NIRCam Deep imaging \citep{Rieke:2023}, taken and publicly released\footnote{All of the \jades\ data used in this paper can be found on the MAST data-base at \url{https://doi.org/10.17909/z2gw-mk31}.} as part of the \textit{JWST} Advanced Deep Extragalactic Survey (JADES: \citealt{Eisenstein:2023}). These data are taken in the GOODS-S field, covering an area of $\sim25~{\rm arcmin}^2$. Specifically, we make use of magnitudes in the F115W, F150W, F200W, F277W, F335M, F356W, F410M, and F444W filters, computed using a Kron parameter of $K = 2.5$, which has been point spread function-convolved to the resolution in the F444W filter, as recommended in the data release. To complete our feature set, we also use photometric redshifts derived using \texttt{EAZY}~\citep{eazy}, as included in the JADES catalogue \citep{Hainline:2024}. Apparent UV magnitudes, $m_{\rm AB}^{1500}$, are computed by fitting a power law ($f_{\lambda}\propto \lambda^{\beta}$) to the three filters nearest to rest-$1500$\AA, selected for each redshift. For a full discussion of the approach as well as comparisons to spectroscopic redshifts the reader is directed to \cite{Hainline:2024} and \cite{Rieke:2023}. Before proceeding, we make the following cuts to reduce our sample:
\begin{itemize}
    \item We require a signal-to-noise ratio (S/N) in all filters redward of F200W to be greater than or equal to 3.
    \item We remove any sources that have been flagged as stars or that are affected by diffraction spikes.
    \item We remove any sources with $M_{\rm UV} \leq -23$ at $z>6$ as these are likely to be dominated by AGN.
    \item We have visually inspected all sources at $z\geq 10$ using ancillary data products from the interactive JADES viewer\footnote{\url{https://jades-survey.github.io/viewer/}} and have removed any spurious sources\footnote{This process removed 14 sources, including low-redshift interlopers and noise-dominated spurious signals.}.
\end{itemize}
Following this process, we are left with a sample of 4,559 galaxies.

For each object, we use the \ac{ILI} pipeline to predict \Nion\ based on observed magnitudes in each filter normalised by $m_{\rm AB}^{1500}$, three colours (F115W-F150W, F150W-F277W, and F277W-F444W), and $m_{\rm AB}^{1500}$. In each case, we account for the model, photometric magnitude, and redshift uncertainties by resampling as described in~\Cref{sec:model}. As an overview, \cref{fig:post_stamps} shows 18 example galaxies from the \jades\ catalogue in redshift bins of $z\in \{6,7,8\}$. Here, we compile RGB images composed of F444W in the red channel, F200W in the green and F090W in the blue. For each object, we also list their observed absolute UV luminosity, predicted value of \Nion\ from the ILI pipeline, and photometric redshift. Galaxies are shown in order of their ionizing photon contributions in each given redshift bin.

Next, in~\cref{fig:muv_nion}, we show the escaped ionizing luminosity of \jades\ galaxies as a function of their observed absolute UV magnitude, coloured by redshift. For comparison, we include spectroscopic measurements from~\cite{Saxena:2024} (\textit{cyan}) as well as \ac{SED} fitted predictions using \texttt{PROSPECTOR} from \cite{Simmonds:2024} (\textit{red}). In both cases, we follow the reported methods of predicting \fesc. In the first case, we use the multivariate model proposed by \cite{Choustikov:2024}, while in the latter we use escape fractions inferred from the absolute UV magnitude ($M_{\rm UV}$), based on the VULCAN simulation \citep{Anderson:2017}. However, we caution that the relation between $M_{\rm UV}$ and \fesc\ has been shown to be very dependent on stellar mass (see Figures 12 and 13 of \citealt{Choustikov:2024}) and is in general not a good predictor for \fesc\ \citep[e.g.][]{Flury:2022b,Saxena:2024,Choustikov:2024}. Both sets of values are included, with intrinsic \nion\ shown as an \textit{arrow} and escaped \nion\ given as \textit{points}.

Here, we can see that there is some correlation between \Nion\ and $M_{\rm UV}$. Galaxies with $M_{\rm UV} < -20$ are rare, but all have large escaped ionizing luminosities ($\dot{n}_{\rm ion} \gtrsim 10^{52}~{\rm photons}/{\rm s}$). We find that UV-dim galaxies with $M_{\rm UV}>-17$ are much more common but have much smaller values of \Nion, with all of these galaxies having $\dot{N}_{\rm ion} \lesssim 10^{53}~{\rm photons}/{\rm s}$. To illustrate the distribution of absolute UV magnitudes, we include a histogram (\textit{top}) comparing the distribution of \jades\ galaxies to those from \sphinx. Beyond confirming that \sphinx\ galaxies are suitable analogues, this shows the sheer number of UV-dim galaxies in our sample. We define an absolute UV magnitude cut at $M_{\rm UV}=-18.5$ (shown as a \textit{dashed} line), which is used to explore whether faint galaxies are the dominant contributors of ionizing photons during the epoch of reionization (\citealt{Finkelstein:2019}, cf.~\citealt{Naidu:2020}).

Figure \ref{fig:nion_z} shows the inferred values of \Nion\ \textit{middle}) and $f_{\rm esc}\xi_{\rm ion}$ (\textit{bottom}) as a function of redshift for all \jades\ galaxies in our sample, along with associated error bars. We colour points by their redshift uncertainty, which in practice was found to be the dominant sources of error in ionizing luminosity. For comparison, we include other observational data from \cite{Saxena:2024} (corrected using the \fesc\ relation of \citealt{Choustikov:2024}; \textit{cyan}) and \cite{Simmonds:2024} (corrected with the $M_{\rm UV}$ relation of \citealt{Anderson:2017}; \textit{red}). Where possible, we also show observational lines of best fit from \cite{Saxena:2024} and \cite{Simmonds:2024c} where we use a fiducial escape fraction of $10\%$. Next, Given we have uncertainties in both \Nion\ and $z$, we use \texttt{ROXY} \citep{roxy}\footnote{\url{https://github.com/DeaglanBartlett/roxy}}, which provides an unbiased linear fit accounting for uncertainties in both $x$ and $y$. We find weak evolutions with redshift, given by:
\begin{align}
     \log_{10}(\dot{N}_{\rm ion}\;/\;[{\rm photons}/{\rm s}]) &= (0.08\pm0.01)z + (51.60\pm0.06), \label{eq:lobf}\\
     \log_{10}(f_{\rm esc}\xi_{\rm ion}\;/\;[{\rm Hz}/{\rm erg}]) & \;= (0.07\pm0.01)z + (24.12\pm0.07),\label{eq:lobf2}
\end{align}
that we also plot (\textit{lime}) with associated $3~\sigma$ uncertainties. We find that this matches our running mean (\textit{blue}) well. Such a slow evolution with $z$ is in agreement with previous works, which suggest little change in \xion\ \citep[e.g.][]{Saxena:2024,Simmonds:2024,Simmonds:2024c} as well as the LyC escape fraction \citep{Mascia:2023}. It is particularly exciting to see how well we agree with \cite{Simmonds:2024c}, given the fact that we are using both using \jades\ observations of GOODS-S. This confirms the validity of our approach. Next, we see that there is a small secondary population present, with \Nion\ lower by about $2~\mathrm{dex}$. These are possibly galaxies with particularly dusty sight lines, for which the model tends to struggle and under-estimates \Nion. In contrast, they may be a population of `remnant leakers' \citep{Katz:2023b} with large escape fractions but low ionizing photon production rates. A discussion of these systems is also given in \cite{Simmonds:2024c}. 

Finally, it is interesting to note that the inclusion of uncertainties in the photometric redshift had the primary effect of increasing the slopes of Equations \ref{eq:lobf} and \ref{eq:lobf2}, which are particularly felt at higher redshifts. This has a similar effect to that seen in Figure 8 of \cite{Simmonds:2024c}, where they find that spectroscopic samples (where there is effectively zero uncertainty in redshift) consistently produced steeper slopes with redshift for both \Nion\ and $f_{\rm esc}\xi_{\rm ion}$. Finally, we note that the outliers described above did not affect the lines of best fit given above.

Next, we use the intrinsic ionizing luminosities ($\dot{N}_{\rm ion,int}^{\rm R22}$) and LyC escape fractions ($f_{\rm esc}^{\rm R22}$) from the \sphinx\ simulation to compute average values for each redshift bin. In doing so, we show the average of the product of these two quantities ($\langle f_{\rm esc}^{\rm R22}\dot{N}_{\rm ion,int}^{\rm R22}\rangle$, representing the ILI approach) as well as the product of their respective averages ($\langle f_{\rm esc}^{\rm R22}\rangle \langle\dot{N}_{\rm ion,int}^{\rm R22} \rangle$, representing the use of population-averaged statistics) in \textit{gold}. We find that in general these two values do not agree, with the latter method over-predicting the average escaped ionizing luminosity by $0.5~$dex toward the end of reionization. This emphasizes the fact that it is the angle-averaged product of these two quantities which is important to measure in order to accurately investigate galaxy contributions to reionization. 

Finally, we can also see the fundamental UV magnitude limit derived from the JADES NIRCam depths~\citep{Eisenstein:2023}. This leads to a reduction in the number of sources with redshift, as \textit{JWST} is able to see fewer sources with the given S/N in each filter. We note that in practice, the trend seen in~\cref{fig:nion_z} remains fairly unchanged with respect to signal-to-noise cuts.

\subsection{Implication for Reionization in GOODS-S}\label{sec:follow-up}

Now that we have predictions for the ionizing luminosity of a large number of galaxies in the GOODS-S field, we can reconstruct a reionization history for the survey volume. To do this, we sum the ionizing luminosity contributions of all galaxies in each redshift bin, while also integrating the comoving volume of each bin, as follows:
\begin{equation}
    \dot{n}_{\rm ion}(z)
    =
    \rho_{\rm UV}(z) \xi_{\rm ion}(z) f_{\rm esc} = \frac{\sum_{z-\delta z}^{z+\delta z}\dot{N}_{\rm ion}(z)}{\int_{z-\delta z}^{z+\delta z} \dd V(z)}.
\end{equation}
This tells us how many ionizing photons are being emitted by galaxies per $\mathrm{Mpc}^3$ in a given redshift bin. This value can then be compared to various models of reionization. It is instructive to use Equation 26 from~\cite{Madau:1999}:
\begin{equation}
    \label{eq:madau1}
    \dot{n}_{\rm ion}
    =
    (10^{51.2} [{\rm photons}/{\rm s}/{\rm Mpc}^3]) ~ C \bigg{(}\frac{1 + z}{6}\bigg{)}^3 \bigg{(}\frac{\Omega_{\rm b} h_{50}^2}{0.08}\bigg{)}^2,
\end{equation}
where $\Omega_{\rm b}$ is the baryonic density fraction of the Universe and $C$ is the ionized hydrogen clumping factor, accounting for the fact that baryons are not uniformly distributed through the \ac{IGM}. In particular, this model depends on a time-dependent clumping factor that is typically calibrated with large-scale simulations (e.g.~\citealt{So:2014}, see also~\citealt{Gnedin:2022} for a review).

\cref{fig:nion_goodss} shows the integrated redshift evolution of \nion\ for all galaxies in our sample, as compared to the theoretical models from~\cite{Finkelstein:2019,Kulkarni:2019}, Bayesian-inferred history from~\citep{Mason:2019} as well as various observational data~\citep{Rinaldi:2023,Mascia:2023b,Simmonds:2024,Simmonds:2024c}. Finally, we include curves showing the number of ionizing photons required to ionize the neutral \ac{IGM} for various clumping factors $C \in \{1, 3, 10\}$ given by~\cref{eq:madau1}. We find that our data is consistent with all of the observations, and predicted histories for the evolution of ionizing photon sources. It is also interesting to explore the question of whether reionization is driven by a small number of UV-bright sources or by a large number of UV-dim sources. To test this, we make a further cut in our data, computing \nion\ for all galaxies in our sample with $M_{\rm UV} < -18.5$ (\textit{magenta}) and $M_{\rm UV} \geq -18.5$ (\textit{purple}), accounting for the two groups respectively. We find that at late times ($z\lesssim 8$) the cohort of UV-dimmer galaxies (that account for $80\%$ of the population) release more ionizing photons into the IGM overall, agreeing with previous work \citep[e.g.][]{Finkelstein:2019}. It is difficult to constrain the two groups' relative importance beyond this redshift due to the difficulty in observing dim galaxies with such a selection function at these distances.

However, there are several key points to discuss. The first is that, as noted previously, our model does not specifically include \ac{AGN}. Therefore, while we do include \ac{AGN} hosts as sources (as we do not make any AGN-related selection cuts apart from removing exceptionally bright sources), our model does not account for any changes in the production or escape of ionized photons induced by the presence of an \ac{AGN}~\citep[e.g.][]{Grazian:2018}. Therefore, we do not observe, for instance, the late-time bump in ionizing luminosity that \ac{AGN} cause~\citep{Kulkarni:2019b,Dayal:2020,Trebitsch:2021}. The second is that at the highest redshifts, our prediction of \nion\ becomes under-estimated due to the UV magnitude limit imposed by \jades\ being a flux-limited survey, thus effectively reducing the completeness of the sample at $z \gtrsim 8$ (see also the discussion in~\citealt{Robertson:2023}).

\begin{figure*}
    \includegraphics{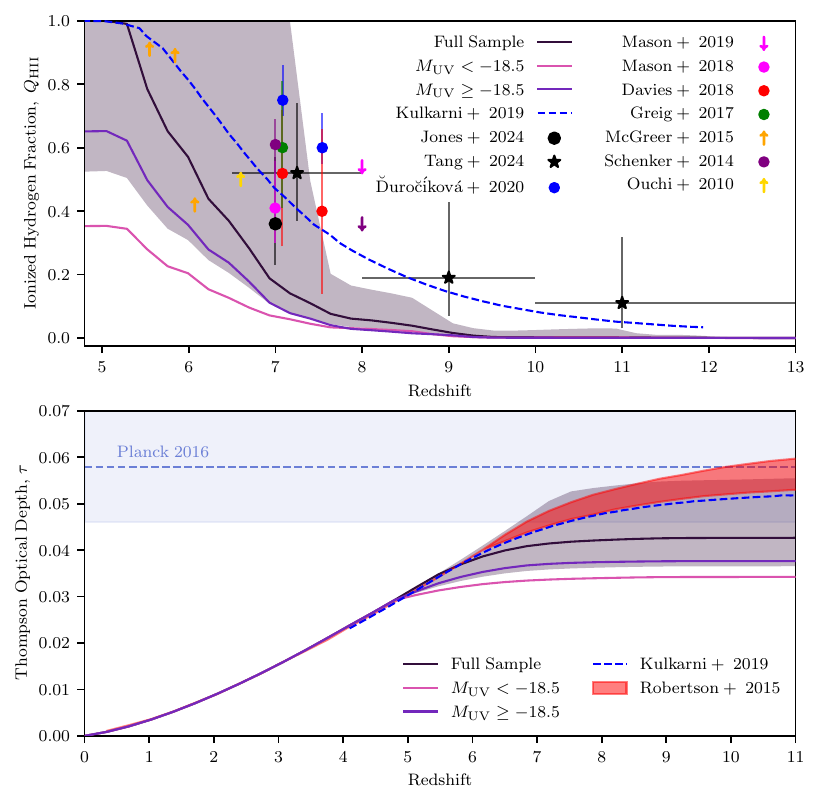}
    \caption{Evolution of the volume-averaged ionized fraction of hydrogen (\textit{top}) as well as the Thompson optical depth (\textit{bottom}) as a function of redshift. We include curves for our full sample (\textit{black}) as well as for only UV-bright galaxies ($M_{\rm UV} < -18.5$; \textit{magenta}) and only the UV-dim galaxies ($M_{\rm UV} \geq -18.5$; \textit{purple}). For the full sample, we also include uncertainties computed by resampling both photometric and model uncertainties (\textit{dark purple shaded region}). UV-bright galaxies are particularly important at high redshift ($z\gtrsim8$), but can only reionize 35\% of the volume by themselves, despite accounting for the brightest 20\% of the sample. The large number of remaining UV-dim galaxies dominate at lower redshifts, reionizing $85\%$ of the survey volume. Thus, neither group are solely responsible, but together are able to drive reionization to completion by $z\sim5.3$. For comparison to $Q_{\rm HII}$, we include simulation results from \protect\cite{Kulkarni:2019} as well as a number of observational results \protect\citep{Ouchi:2010,Schenker:2014,McGreer:2015,Greig:2017,Davies:2018,Mason:2018,Mason:2019b,Durovcikova:2020,Jones:2024,Tang:2024}. Likewise for $\tau$ we compare to results from \protect\cite{Robertson:2015,Kulkarni:2019} as well as constraints from \textit{Planck} \protect\citep{Planck:2016}.}
    \label{fig:ionized_frac}
\end{figure*}

Now, we aim to use this sample to produce an explicit reionization history, tracing the evolution of the ionized fraction ($Q_{\rm HII}$) based on only our sample of galaxies. To do this, we make use of the modified ``reionization equation'' of \cite{Madau:2017}:
\begin{equation}\label{eq:madau2}
\frac{d Q_{\rm HII}}{dt} = \frac{\dot{n}_{\rm ion}}{\langle n_{\rm H}\rangle(1 + \langle \kappa_{\nu_L}^{\rm LLS}\rangle / \langle\kappa_{\nu_L}^{\rm IGM}\rangle)} - \frac{Q_{\rm HII}}{\overline{t}_{\rm rec}},
\end{equation}
where $\langle n_{\rm H}\rangle = 1.9\times10^{-7}~{\rm cm}^{-3}$ is the comoving number density of hydrogen in the \ac{IGM} \citep{Gnedin:2022} and $\langle\kappa_{\nu_L}^{\rm LLS}\rangle$ and $\langle\kappa_{\nu_L}^{\rm IGM}\rangle)$ are absorption coefficients due to high-density clumps known as Lyman-limit systems~\citep{Crighton:2019,Becker:2021,Zhu:2023,Georgiev:2024} as well as the \ac{IGM} itself. This term is proportional to $1 - Q_{\rm HII}$ and becomes important as ionized bubbles begin to merge and overlap (at $z\sim6$), accounting for the presence of optically thick absorbers that ensure that the mean-free path of LyC photons remains small once overlap begins to occur \citep{Gnedin:2006,Furlanetto:2009,Worseck:2014}. The ratio of these two quantities is given as a function of redshift in Equation 32 of~\cite{Madau:2017} and is taken as 0 for $z>6$. Finally, $\overline{t}_{\rm rec}$ is an ``effective'' recombination timescale in the \ac{IGM}. For our purposes, we use the following fitting formula:
\begin{equation}\label{eq:rec}
    \overline{t}_{\rm rec} = 2.3\bigg{(} \frac{1 + z}{6}\bigg{)}^{-4.35}~{\rm Gyr},
\end{equation}
based on analysis of a radiation hydrodynamical simulation by~\cite{So:2014}. We choose this expression because it does not require an estimate of the clumping factor $C$, although much work has been carried out to estimate redshift-dependent values of $C$ using cosmological hydrodynamic simulations \citep{Kohler:2007,Pawlik:2009,Finlator:2012,Shull:2012,So:2014,Kaurov:2014}. We note, however, that there is evidence for a large galaxy over-density in GOODS-S at $z\sim 5.4$ \citep{Helton:2024}, which may further stress the effectiveness of this approximation at low redshifts, towards the end of reionization. In fact, as expected, we also find a slight bump in \nion\ (Figure \ref{fig:nion_goodss}) at this redshift.

Another key quantity to compute is the Thompson optical depth to the microwave background, $\tau$. This can be computed as \citep{Kuhlen:2012,Robertson:2015,Robertson:2022}:
\begin{equation}\label{eq:tau}
    \tau(z) = c\sigma_T \langle n_{\rm H}\rangle \int_0^z dz'~ \frac{(1+z')^2}{H(z')}\bigg{[}1 + \frac{\eta Y}{4X}\bigg{]} Q_{\rm HII}(z'),
\end{equation}
where $c$ is the speed of light, $\sigma_T$ is the Thompson cross section, and we assume that helium is fully ionized ($\eta=2$) at redshifts $z < 4$ and singly ionized ($\eta=1$) before this.

Using these expressions, as well as our results from~\cref{fig:nion_goodss}, in~\cref{fig:ionized_frac} we show the computed evolution histories and associated uncertainties for $Q_{\rm HII}$\footnote{Due to the flux limits of the survey, we solve Equation \ref{eq:madau2} from $Q_{\rm HII} = 0$ at $z = 13$. We also artificially set $Q_{\rm HII} = 1$ once reionization is complete. This is due to the fact that~\cref{eq:madau2} is only valid until a given patch is nearing complete reionization. The interested reader is directed to discussions in \cite{Robertson:2013,Madau:2017,Gnedin:2022}.} (\textit{top}) and $\tau$ (\textit{bottom}). In the case of $Q_{\rm HII}$, we compare to results from \cite{Kulkarni:2019} as well as observational constraints \citep{Ouchi:2010,Schenker:2014,McGreer:2015,Greig:2017,Davies:2018,Mason:2018,Mason:2019b,Durovcikova:2020,Jones:2024,Tang:2024}. For $\tau$, we compare to results from \cite{Kulkarni:2019,Robertson:2015} as well as constraints from \cite{Planck:2016}.

We find that the galaxies considered within this sample are able to complete reionization within the GOODS-S region by $z\sim5.3$. Furthermore, the rapid evolution in $Q_{\rm HII}$ for the full sample begins very late, being only $\sim20\%$ complete at $z = 7$, in agreement with various observational probes favouring a relatively late reionization \citep{Ouchi:2010,Schroeder:2013,Schenker:2014,Onorbe:2017,Banados:2018,Villasenor:2022} as well as \textit{Planck} \citep{Planck:2016}. Of particular note, are comparisons with the other indirect \textit{JWST}-based approaches of \cite{Jones:2024} and \cite{Tang:2024}. Both use Ly$\alpha$ emitters to constrain the evolution of the IGM neutral fraction, with the key difference being that the former uses observations in GOODS-S and GOODS-N while the latter uses a more varied number of fields (GOODS-S, GOODS-N, Abell-2744, EGS). Therefore, the fact that we are consistent with the completeness-corrected approach of \cite{Jones:2024}\footnote{It is important to note that this method is based around the work of \cite{Dijkstra:2011}, which implicitly assumes that reionization concludes at $z = 6$. This may have the effect of biasing their value of $Q_{\rm HII}$ upwards with respect to our line.} using a very similar JADES sample is another confirmation of the validity of our approach. On the other hand, disagreements with \cite{Tang:2024} highlight the effect of cosmic variance, confirming the fact that our results are only applicable to GOODS-S.

Finally, as before, we repeat this calculation for the UV-bright and UV-dim galaxies defined in~\cref{fig:muv_nion}. Here, we find that UV-bright galaxies are only able to reionize $\sim35\%$ of the volume by themselves, despite accounting for the brightest 20\% of the population. On the other hand, the larger number of UV-dim galaxies become completely dominant at $z<7.5$, managing to ionize $\sim65\%$ of the volume by themselves. As such, we conclude that neither group of sources is able to reionize the Universe on time solely by themselves, but that the complete set of star-forming galaxies are able to complete reionization without the help of \ac{AGN} or more exotic sources of ionizing photons \citep{Furlanetto:2008,Robertson:2015,Liu:2016,Kulkarni:2019b,Dayal:2020,Ma:2021,Saxena:2021,Trebitsch:2021,Trebitsch:2023}.

An important caveat is the fact that in this analysis we are only integrating so far down the UV luminosity function, owing to the flux limited sample of JADES\footnote{For a complete discussion, the reader is directed to \cite{Robertson:2023}.} \citep{Eisenstein:2023} as well as our selection function. In doing so, we are not completely sampling galaxies at fainter magnitudes (particularly at $z\gtrsim7$), with no meaningful representation at $M_{\rm UV}\geq-15$. In turn, these sources may have comparable \nion\ contributions, despite the potential turnover at the faint end of the UV luminosity function \citep{Bouwens:2022,Williams:2024}. For example, recent work by \cite{Wu:2024} suggests that for a constant \fesc, dwarf galaxies with $M_{\rm UV} > -14$ might contribute $\approx$40-60\% of the ionizing photon budget at $z>7$, reducing to $\approx$20\% at $z=6$, highlighting the need to account for these objects. In practise, including these sources will increase \nion\ (particularly at higher redshifts), thus particularly modifying the intermediate reionization history and making it conclude slightly earlier, potentially in line with other observational constraints. As such, it would be interesting to repeat this exercise with other deep surveys (e.g. JADES Origin Field; \citealt{JOF}, NGDEEP; \citealt{NGDEEP}, GLASS; \citealt{GLASS}), wider surveys (e.g. CEERS; \citealt{CEERS}, PRIMER; \citealt{PRIMER}, COSMOS-WEB; \citealt{COSMOS-WEB}) and particularly those which are targeted at lensing clusters which can push to even fainter UV luminosities (e.g. UNCOVER; \citealt{uncover}). We leave such explorations to future work, though note that our model can also be trained on other sets of \textit{JWST} filters and is therefore suitable for these applications.

In the case of the Thompson optical depth, we recover a redshift evolution in agreement with previous results from \cite{Robertson:2015,Kulkarni:2019} up to $z\sim6$. However, at redshifts beyond this, we similarly find that the reduced number of sources in our sample at higher redshift leads to a value of $\tau =0.043$, falling below the constraints from \textit{Planck} \citep{Planck:2016}. In agreement with the evolution of $Q_{\rm HII}$, the majority of optical depth evolution is driven by UV-dim galaxies, confirming their importance.

It is important to note that we do not suggest that the curve shown in~\cref{fig:ionized_frac} is the definitive history of reionization in GOODS-S, particularly given that we do not have a complete sample by definition (see the selection described in~\Cref{sec:jades}). Instead, the purpose of this work has been to show that galaxy properties such as (but not limited to\footnote{In principal such a method (using \ac{ILI} applied to photometry) can be leveraged to predict any galaxy property included in the \sphinx\ public data release.}) \Nion\ can be self-consistently derived from photometry. It is, however, particularly interesting that the sample studied here is able to drive reionization to completion on a realistic time-scale, leaving space for ever dimmer galaxies to make their mark. In conclusion, this work further accentuates the fact that while \textit{JWST} has certainly ushered in a new era for the study of reionization, it is necessary to use deep surveys with well-defined selection functions and self-consistent models to build a complete picture of cosmic dawn.

\section{Caveats}\label{sec:caveats}

Like any new method, the approach that we have presented comes with a fair share of assumptions and caveats, primarily due to the fact that our model has been trained on a simulation. Therefore, it can be instructive to highlight these avenues for future improvement:

\textbf{\textit{Realism of \sphinx:}} As has been discussed extensively, simulations such as \sphinx\ are reliant on an array of sub-grid prescriptions for star-formation, feedback and ISM processes, variations in which can have significant effects on galaxy properties and evolution. While these do not accurately capture the microphysics which must be occurring, \sphinx\ still represents the present-day state-of-the-art simulation owing to its combination of size and resolution enabling the simulation of a statistical sample of galaxies with multi-phase \acp{ISM} (see discussion in Section 4 of \citealt{spdrv1}). Nevertheless, \sphinx\ has been shown to reproduce a large number of observed properties relevant to the production and escape of ionizing radiation, such as relationships between LyC escape fraction and spectral properties \citep{Choustikov:2024}, UV luminosity functions \citep{Rosdahl:22}, Ly$\alpha$ luminosity functions \citep{Garel:2021}, Ly$\alpha$ spectral properties \citep{Choustikov:2024b} as well as others \citep{spdrv1}. As a result, while it is always push to better-resolved simulations with more physics, all of this fills us with confidence that galaxies in \sphinx\ act as good analogues for galaxies observed during the EoR. Taken at their absolute worst, \sphinx\ galaxies represent complex, 3D photoionization models with realistic SFHs and ISM structure, replaced only by future simulations accounting for non-equilibrium metal abundances and thermochemistry  \citep{Katz:2022d,Katz:2022c,Katz:2024b}.

\textbf{\textit{Lack of AGN:}} The largest caveat to our work is that we have not accounted for the presence of AGN, both through the contribution of feedback to clearing channels for LyC escape or the hard spectra of ionizing radiation produced by the inner regions of accretion disks. This is not expected to be an issue, as spectroscopic follow up by \jades\ found that roughly $\sim5\%$ of galaxies hosted broad-line AGN \citep{Maiolino:2023}, suggesting a low AGN fraction among the sources we have considered. Nevertheless, the model derived in this work can only account for the star-forming contributions of a given source. While AGN contributions are expected to not be significant \citep{Kulkarni:2019b,Dayal:2020,Ma:2021,Trebitsch:2021,Trebitsch:2023}, we conclude by confirming that the reionization history shown in Figure \ref{fig:ionized_frac} can only represent a lower-bound, especially accounting for dim galaxies below the detectability limit of JADES.

\textbf{\textit{Photometric Redshifts:}} Unsurprisingly, the approach taken in this work relies on the accuracy of photometric redshifts estimated using the JADES filter-set, which themselves are depth, filter-set and selection dependent. While this is always an area that can be improved (\citealt{Newman:2022}, see also Figure 4 of \citealt{Bouwens:2023}), the accuracy of these measurements has proven surprisingly good \citep[e.g.][]{Hainline:2024}. In the end, photometric catalogs such as those used in this work represent the only possible present-day approach given the cost of spectra \citep{Bunker:2024}. Nevertheless, while we consistently account for them in our error estimation, photometric uncertainty remains the dominant source of error for the ILI model.

\textbf{\textit{Cosmic Variance:}} Given that the GOODS-S survey is taken in a small area of sky ($\sim25$ square arcminutes), it is important to consider the effect of cosmic variance on our results \citep[e.g.][]{Kragh-Jerspersen:2024}. For instance, the presence of overdensities and rare objects can all contribute to a skewed reionization history. Here, while it is important to note the consistency between our results and those of different approaches applied to observations in the same survey \citep{Simmonds:2024c,Jones:2024}, any direct conclusions we make are only immediately applicable to the GOODS-S survey volume.

Despite these sources of uncertainty, we conclude that while our approach can be improved with more advanced modeling and application to a more extensive data-set, it still represents the first approach of its kind: using simulation-trained ILI to directly infer the ionizing contributions of photometric galaxies during the Epoch of Reionization.

\hspace{2cm}

\section{Conclusions}\label{sec:conclusion}

We present \texttt{PHOTONIOn}\footnote{\url{https://github.com/Chousti/photonion.git}}: an \acl{ILI} (\ac{ILI}) model based on the \ltuili\ pipeline \citep{Ho:2024} to predict the angle-averaged escaped ionizing luminosity, \Nion, of Epoch of Reionization galaxies based on observed photometric magnitudes and redshifts. Trained on 13,800 mock dust-attenuated photometric line-of-sight measurements of \textit{JWST} analogues from the \sphinx\ simulation \citep{spdrv1}, this model has been validated and shown to perform better than estimates computed using standard \ac{SED}-fitting techniques which typically rely on an independent method to estimate the LyC escape fraction (\fesc), including better performance across multiple sight-lines for the same object. One of the key novelties of our model compared to previous analyses is that rather than treating the ionizing photon production rate and LyC escape fraction as separate quantities, they are inferred together. Hence, our method does not require a separate prescription for \fesc, or for any dust-correction, as these relations have been self-consistently learnt by the model.

This ILI model was then deployed on a sample of 4,559 photometrically observed galaxies in the GOODS-S field as part of the \jades\ programme \citep{Eisenstein:2023}, allowing us to explore the redshift evolution of \Nion, the number density of ionizing photons released into the \acl{IGM} (\ac{IGM}), \nion, as well as the volume-averaged ionized fraction of hydrogen, $Q_{\rm HII}$.

Our conclusions are as follows:

\begin{itemize}
    \item We show that \ac{ILI}-based approaches trained on sophisticated cosmological simulations are capable of inferring non-trivial observables depending on small-scale ISM physics, confirming an interesting avenue for inference pipelines of the future.
    \item \texttt{PHOTONIOn} is capable of accurately inferring \Nion from photometry, while also producing self-consistent uncertainties. Additionally, it is orders of magnitude faster than traditional \ac{SED}-fitting methods, allowing for easy application to large datasets.
    \item When compared to a standard SED-fitting approach, it was found that the dominant source of uncertainty is due to the models used to infer \fesc, even if access to spectroscopic data is assumed. While standard prescriptions often over-predict the contributions of LyC-dim galaxies, \texttt{PHOTONIOn} is able to overcome this difficulty and performs more consistently due to the fact that it has implicitly learnt a relation between galaxy colours and \fesc.
    \item Methods of estimating \fesc\ based on the UV-continuum slope, $\beta$, perform poorly when applied to galaxies with non-negligible contributions from the nebular continuum. This further emphasises the need to account for the nebular continuum when considering the contributions of galaxies to reionization \citep[see also][]{Katz:2024,Saxena:2024b}.
    \item For a sample of 4,559 \jades\ photometric galaxies, \Nion\ and $f_{\rm esc}\xi_{\rm ion}$ both evolve slowly with redshift, as: $\log_{10}(\dot{N}_{\rm ion}) = (0.08\pm0.01)z + (51.60\pm0.06)$ and $\log_{10}(f_{\rm esc}\xi_{\rm ion}) = (0.07\pm0.01)z + (24.12\pm0.07)$.
    \item Star-forming galaxies observed within this sample are capable of producing a reionization history that begins late and completes at $z\sim 5.3$.
    \item UV-dim galaxies (with $M_{\rm UV}\geq -18.5$, accounting for $80\%$ of the sample) are able to reionize the majority of the survey volume, contributing more to reionization both in terms of the ionized Hydrogen fraction ($Q_{\rm HII}$) and Thompson optical depth ($\tau$) as compared to UV-bright galaxies (with $M_{\rm UV}< -18.5$, $20\%$ of the sample). Thus, we find that neither subgroup is capable of driving reionization by themselves but faint galaxies appear to be crucial.
\end{itemize}

We have utilised the synergy of photometric \textit{JWST} observations and cosmological radiation hydrodynamic simulations with a resolved multi-phase \acl{ISM} to build an inference pipeline for the luminosity of ionizing photons released into the \ac{IGM} by galaxies during the Epoch of Reionization. Beyond providing valuable insight into the contributions of star-forming galaxies to the evolution of reionization, this work further highlights the necessity for observers and simulators to work together as we continue to explore the cosmic dawn.

\section*{Acknowledgements}

We would like to thank the anonymous referee for inspiring changes which improved the quality and depth of this work. We thank Matthew Ho, Brant Robertson, and Joel Leja for insightful discussions. The authors thank Jonathan Patterson for smoothly running the Glamdring Cluster hosted by the University of Oxford, where part of the data processing was performed.

N.C. acknowledges support from the Science and Technology Facilities Council (STFC) for a Ph.D. studentship, as well as the hospitality of Princeton's Department of Astrophysical Sciences where part of this work was completed. RS acknowledges financial support from STFC Grant No. ST/X508664/1. AS acknowledges funding from the ``FirstGalaxies'' Advanced Grant from the European Research Council (ERC) under the European Union's Horizon 2020 research and innovation programme (Grant agreement No. 789056)

This work used the DiRAC@Durham facility managed by the Institute for Computational Cosmology on behalf of the STFC DiRAC HPC Facility (www.dirac.ac.uk). The equipment was funded by BEIS capital funding via STFC capital grants ST/P002293/1, ST/R002371/1 and ST/S002502/1, Durham University and STFC operations grant ST/R000832/1. DiRAC is part of the National e-Infrastructure.This work was performed using the DiRAC Data Intensive service at Leicester, operated by the University of Leicester IT Services, which forms part of the STFC DiRAC HPC Facility (www.dirac.ac.uk). The equipment was funded by BEIS capital funding via STFC capital grants ST/K000373/1 and ST/R002363/1 and STFC DiRAC Operations grant ST/R001014/1. DiRAC is part of the National e-Infrastructure.

\section*{Author Contributions}
The main roles of the authors were, using the CRediT (Contribution Roles Taxonomy) system\footnote{\url{https://authorservices.wiley.com/author-resources/Journal-Authors/open-access/credit.html}}:

\textbf{Nicholas Choustikov}: Conceptualization; Formal analysis; Methodology; Software; Visualisation; Writing - original draft. \textbf{Richard Stiskalek}: Conceptualization; Methodology; Software; Writing - original draft. \textbf{Aayush Saxena}: Conceptualization; Writing - review and editing. \textbf{Harley Katz}: Conceptualization; Writing - review and editing. \textbf{Julien Devriendt}: Supervision; Resources; Writing - review and editing. \textbf{Adrianne Slyz}: Supervision; Resources; Writing - review and editing.


\section*{Data Availability}
The code behind \texttt{PHOTONIOn} and models used in this work are available at \url{https://github.com/Chousti/photonion.git}. The \sphinx\ data used in this work is available as part of the SPHINX Public Data Release v1 (SPDRv1,~\citealt{spdrv1}), available at \url{https://github.com/HarleyKatz/SPHINX-20-data}. The \jades\ photometric catalogue used in this work is available at \url{https://archive.stsci.edu/hlsp/jades}. All other data will be shared upon reasonable request to the corresponding author.


\bibliographystyle{mnras}
\bibliography{References.bib}

\appendix
\section{Validating the Model on \sphinx}\label{sec:validate}

Here, we proceed to complete a variety of benchmark tests on the model described in Section \ref{sec:model}.

\begin{figure}
    \includegraphics{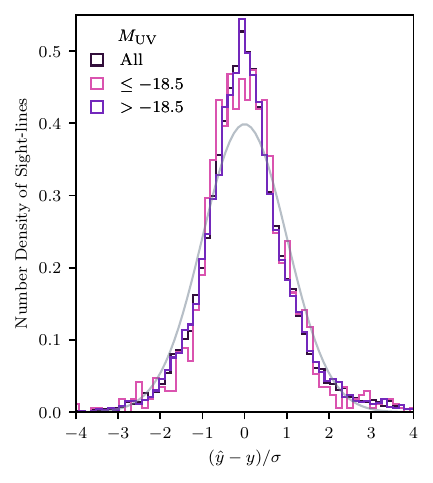}
    \caption{Standardised residuals (as defined in Eq.~\protect\ref{eq:renormalisation} for \ac{ILI} predictions, for the full sample of (\textit{black}), UV-bright ($M_{\rm UV} \leq -18.5$; \textit{magenta}), and UV-dim ($M_{\rm UV} > -18.5$; \textit{purple}) sight-lines. We include the means (\textit{bold}) and standard deviations (\textit{dashed}), as well as a standard Gaussian ($\mathcal{G}(0, 1)$; \textit{gray}) for comparison}
    \label{fig:calibration}
\end{figure}

First, in order to further confirm that the uncertainties produced by the model are self-consistent, in~\cref{fig:calibration} we show histograms of the standardised residuals given by:
\eq{
\label{eq:renormalisation}
x \equiv \frac{\dot{N}_{\rm ion}^{\rm predicted} - \dot{N}_{\rm ion}^{\rm true}}{ \langle {\rm unc}(\dot{N}_{\rm ion}^{\rm predicted})\rangle},
}
where $\langle{\rm unc}(\dot{N}_{\rm ion}^{\rm predicted})\rangle$ is the average of the asymmetric $1\sigma$ uncertainties of the \ac{ILI} posterior. We include histograms for the full sample (\textit{black}) as well as for the observed UV-bright ($M_{\rm UV} \leq -18.5$; \textit{magenta}) and UV-dim ($M_{\rm UV} > -18.5$; \textit{purple}) sight-lines of \sphinx\ galaxies. For completeness, we also show the standard Gaussian distribution, $\mathcal{G}(0, 1)$, as a comparison. We find that in all three cases our \ac{ILI} model performs very well, without significant outliers.

\begin{figure}
    \includegraphics{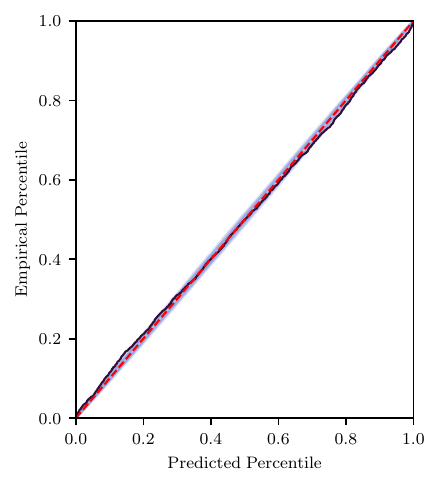}
    \caption{Probability integral transform diagnostic for the ILI model, quantifying the proportion of posterior samples that are below the true value.}
    \label{fig:PIT}
\end{figure}

To further reinforce this point, we also inspect the \acl{PIT} (\acs{PIT},~\citealt{cook_2004}) diagnostic, shown in~\cref{fig:PIT}. The \ac{PIT} quantifies the proportion of posterior samples $\boldsymbol{\theta}$ that are below the true value. If the distribution of \ac{PIT} values is uniformly distributed, then the predicted posterior distributions are consistent with the true values~\citep{Zhao_2021}. The \ac{PIT} distribution is typically assessed on a percentile-percentile plot, which compares the cumulative density function of \ac{PIT} values to that of a uniform random variable. If the learnt posterior is well-calibrated, then the two cumulative density functions should agree. If not, the \ac{PIT} plot is a useful probe of a global bias or over- and under-dispersion. We verify that the test-set \ac{PIT} distribution of our \ac{ILI} model predicting $\log_{10} \dot{N}_{\rm ion}$ passes this test. While the results of Figure \ref{fig:calibration} suggest that the residuals are not in agreement with a standard Gaussian, the fact that the errors are well calibrated (Figure \ref{fig:PIT}) signals that the predicted posteriors are non-Gaussian.

Lastly, we also verify that the predictions of our \ac{ILI} model agree with those of an \acl{ET} regressor (\acs{ET},~\citealt{extra-trees}) as implemented in \texttt{scikit-learn}~\citep{scikit-learn}.
While this is not a validation of the \ac{ILI} pipeline, it is nevertheless a useful sanity check. We similarly optimize the hyper-parameters of the \ac{ET} model and find that the maximum posterior \ac{ILI} and \ac{ET} predictions are correlated with a Spearman correlation coefficient of $0.98$ and that with respect to the true values the \ac{ILI} model marginally outperforms the \ac{ET}, while also providing self-consistent uncertainties.


\bsp	
\label{lastpage}
\end{document}